%% file: shadowing.tex
\ifx\mnmacrosloaded\undefined \input mn\fi
\input epsf
\input psfig.sty
\epsfverbosetrue
\pageoffset{-2.5pc}{0pc}

\begintopmatter

\title{Contaminants in ATCA baselines with shadowing: a case study 
of cross talk in short-spacing interferometers}

\author{Ravi Subrahmanyan$^{1}$, Avinash A. Deshpande$^{1,2,3}$}

\affiliation{$^1$ Australia Telescope National Facility, CSIRO, 
	          Locked bag 194, Narrabri, NSW 2390, Australia}
\vskip 0.1 truecm
\affiliation{$^2$ Raman Research Institute, Sadashivanagar, 
             Bangalore 560 080, India}
\vskip 0.1 truecm
\affiliation{$^3$ Arecibo Observatory, NAIC, HC3 53995, 
             Arecibo, PR 00612, USA}

\shortauthor{R. Subrahmanyan \& A. A. Deshpande}

\shorttitle{Cross talk at the ATCA}

\abstract{
	Interferometric telescopes made of close-packed 
	antenna elements are an important tool for imaging extended
	radio sources, specifically structures that have angular
	sizes comparable to or even greater than the 
	FWHM of the beams of the antennas.  They have proved useful in 
        observations of cosmic microwave background anisotropies 
        that require high brightness-sensitivity.
        However, the visibilities measured in baselines formed between
	close antenna elements -- in particular, between shadowed 
	elements -- of Fourier-synthesis arrays are often observed 
	to be corrupted.  We discuss the multiplicative and 
	additive errors affecting such short-baseline interferometers.

	As a case study, we have examined the nature 
	of the spurious correlations between the 
	Cassegrain-type paraboloidal reflectors that are elements
	of the Australia Telescope Compact Array.  In configurations
	with geometric shadowing, the cross talk here appears as an
	additive component.  Analysis of the characteristics of this
	cross talk leads us to believe that when these reflector antennas
	are in a shadowed configuration, the receivers in the 
        antenna pair pick up correlated emission from opposite sides
	of the main reflector surface of the front antenna.
	The slots between the panels that make up the main reflector surface
	provide the pathway for the coupling across the reflector surface. 
	This mode of cross talk may be avoided by constructing the main
	reflectors of short spacing interferometers as continuous
	conducting surfaces.	
	}


\maketitle

\section{Introduction}

The techniques of radio interferometry and Fourier Synthesis arrays
for imaging have developed remarkably since the first synthesis
images were constructed by Ryle, Hewish \& Shakeshaft (1959) and
Earth-Rotation Fourier synthesis was described by Ryle (1962) more than 
four decades ago.  Today, large aperture synthesis arrays of 
precision paraboloidal reflectors like, for example, the Very Large
Array, the Very Long Baseline Array, the 
Australia Telescope Compact Array (ATCA) and the Westerbork Synthesis
Radio Telescope
routinely provide astronomers with visibility measurements with which
high-resolution and high-dynamic-range radio images of celestial objects 
are synthesized.  The quest for extreme angular resolution has
led to the development of space Very Long Baseline 
Interferometry which involves antennas separated by
distances exceeding the Earth diameter.

For years, conventional wisdom argued against using interferometers 
for observations which required high 
surface brightness sensitivity; total-power
measurements with single-dish telescopes were the preferred option.
The reason is that interferometers are 
`correlated-flux sensitive' telescopes
which respond well to discrete `point sources', they do not respond
to the uniform sky background and couple poorly to extended emission
that is `resolved' by the interferometer spacing.
For example, prior to 1990, all successful attempts to detect the 
Sunyaev-Zeldovich Effect (SZE; Sunyaev \& Zeldovich 1972) 
towards clusters of galaxies were made with
single-dish telescopes (see, for example, the compilation in Birkinshaw
(1990)).  However, this conviction has changed in the last
decade: the advantages of interferometry
and Fourier synthesis -- its inherent ability to provide significant
rejection of any contributions from 
the atmosphere and ground and the stability in its
measurements -- are often perceived to outweigh the disadvantages.  
Interferometers formed 
between closely spaced elements provide visibility measurements of
extended structure and, with recent improvements in receiver technology
which allow low-noise amplifiers and compact cryogenics, it is now possible
to construct useful interferometers of small aperture elements and with
short baselines.  Indeed, today we have 
Fourier-synthesis images of the SZE 
towards several tens of clusters (see, for example, 
Jones et al. 1993, Carlstrom et al. 2000).

The quest for high brightness sensitivity and the scientific motivations
for imaging extended low-surface-brightness features in the sky have
pushed telescope designers towards  arrays of small-sized elements in 
ultra-compact configurations.  
An example of such a science goal is the imaging of temperature variations in 
the cosmic microwave  background (CMB): on large angular scales the 
anisotropies are believed to be the
imprint of primordial density perturbations and on smaller angular scales
they are due to the propagation
of the relic photons through the extended gaseous coronae of galaxy clusters
and diffuse hot intergalactic gas.
The signals are extremely weak: the primary
CMB anisotropy is a few 10s of $\mu$K, the polarized CMB
anisotropy is expected to be a few $\mu$K.
These are a tiny fraction of the
telescope system temperature and a very important design consideration is
the control and rejection of spurious systematics. 
These key astrophysical questions are being probed today by a small
but growing number of `special-purpose'
aperture synthesis telescopes which have been built using elements of
small diameter apertures.  
These ground-based telescopes are imaging the CMB anisotropies with
poor angular resolutions/scales of about a degree. 

Interferometric imaging via Earth-rotation 
Fourier synthesis has the added advantage that 
the `fringe filtering' provides
rejection against systematics that do not have the same fringe rate
as that for the celestial signal of interest. Consequently, it is advantageous
if the antenna elements forming the interferometric telescope are not
co-mounted (on a platform) but are independantly mounted.  
The downside of such an approach is that the antenna elements run the risk of
shadowing when the portion of the wavefront incident on an antenna element 
aperture is blocked along the path by parts of the aperture of
another antenna element located adjacent and in front.  
The visibilities measured by an interferometer pair in a shadowed
configuration, or in a situation where the projected element apertures
are close to shadowing, are usually corrupted.  However, it is these
very baselines between apertures with close projected spacings that 
provide the highest sensitivity to large angular scale 
variations in sky surface brightness temperature.

To avoid shadowing in small arrays, 
a co-mounted array design may be preferred.
However, these arrays in which the apertures are fixed to a platform
will be more susceptible to spurious pickups; consequently, 
they entail an investment of considerable effort towards the development
of appropriate calibration strategies to subtract the 
unwanted additive contributions, particularly those local to the
platform (see Subrahmanyan (2002) for an alternate viewpoint).
Making choices in the design of ultra-compact arrays 
requires understanding the trade-offs: an
important aspect is understanding the magnitude and the origin of
spurious contributions that arise between closely spaced antennas.

The next generation radio telescopes, like the Atacama Large Millimeter
Array, Allan Telescope Array and the Square Kilometre Array, which 
are to have large total collecting areas for high sensitivity, are being
designed to be arrays of small elements.  The arrays would have closely
spaced interferometers for wide-field imaging of extended emission. 
Because short-spacing interferometers are important for recovering
low spatial frequencies (sometimes using mosaicing techniques), understanding
the causes for spurious contaminants in the visibility measurements
made using closely spaced elements
is important for designing these next generation arrays.

In this paper, we examine the nature of the spurious signals that are
observed in those short-spacing interferometers of the ATCA 
when there is geometric shadowing.  
We begin in Section~2 by listing the different mechanisms that corrupt
visibility measurements made using short-spacing interferometers.  
Observations of spurious visibilities in baselines between ATCA antennas
in which one element shadows the other is described in Section~3;
later sections describe investigations into the nature of this
unwanted cross-talk.  

\section{The errors in interferometer baselines with geometric shadowing}

In an interferometer that is formed between antenna elements 
which are identical and spaced well apart,
the measured visibility may be described as
an integral over the product of 
(i) the sky brightness distribution,
(ii) the primary beam pattern of the antenna elements and 
(iii) the interferometer (angular) pattern that 
is determined by the projected antenna spacing or baseline.
In the visibility domain, the spatial frequency range that is sampled by the 
measured visibility on a baseline is determined
by the aperture illuminations of the two antenna elements and 
the projected interferometer spacing.
In interferometer arrays, closely spaced antenna elements that
are independently mounted would shadow each other when the projected
baseline is shorter than the average physical extent of the two apertures 
viewed along the baseline; a situation encountered when observing
in particular azimuth/elevation zones.  
Shadowing causes the measured visibilities to differ 
from that expected based on the above integral
and the calibration of the measurements may suffer.

Some of the possible multiplicative errors in these
visibility measurements are listed below; a few of these
assume that the array elements are reflector-type antennas.
\beginlist
\item (i) The presence of the shadowing antenna element in front 
of the shadowed antenna distorts the far-field antenna radiation pattern 
corresponding to the shadowed antenna.
The blocking of the aperture of the shadowed antenna reduces
its effective collecting area.  If we consider baselines between
a shadowed antenna and any other array element that is unblocked, 
these interferometers may be considered to be formed between a pair 
non-identical apertures.  The effective primary beam pattern in
an interferometer is the geometric mean of the individual element
antenna patterns and, therefore, in the geometric optics (GO) approximation 
the effective primary beam for the case where the aperture of 
one element is shadowed may be computed 
from (the Fourier transform of) the convolution of
the illuminations corresponding to the partially blocked aperture 
with an unblocked aperture.  This convolution defines the weighting, in 
the visibility domain, for the integral that yields the visibility
measurement on the baseline between the elements.
\item (ii) The `effective' baseline to the shadowed antenna 
is modified and may vary also
with the extent of shadowing: the interferometer (or array) pattern
for baselines to the shadowed antenna will vary with the shadowing
geometry.  
\item (iii) The system temperature of the shadowed antenna may be elevated
because there may be absorbing elements in the backup structure of
the antenna in front (which emit with a brightness corresponding to
the ambient temperature) and/or the shadowed antenna
may now pick up ground emission reflected off the structure of the
antenna in front.
\item (iv) Diffraction and scattering will occur at the edge of the antenna 
in front and these imply that
the wavefront is no longer planar as it arrives at the 
aperture of the shadowed
antenna.  This necessitates a physical optics (PO) treatment 
of the blocking.
\endlist

The above list was of multiplicative distortions/errors in the measured
visibility.  There would also be spurious additive errors, some of 
these are listed below.
\beginlist
\item (i) The coupling between the aperture fields of 
two nearby apertures (see, for example, Padin et al. 2000) would
result in receiver noise, which radiates out into
the aperture of one element, to be coupled to the aperture of any 
adjacent element.
Receiver noise may also leak between close antennas as a result of
coupling pathways created by scattering off structural elements like,
for example, sub-reflector support structure.  Coupling paths may also
exist as a result of transmission through the main reflector surface
if this surface is not perfectly reflecting/opaque 
at the observing frequency. The main
reflector may be porous because the surface might be a mesh, or because the
reflector panels might have holes to reduce wind loading, or because 
the main reflector might be constructed from solid panels and there might
be slots between the panels.
\item (ii) The short baseline may respond to the 
near-field atmospheric emission in overlapping near-field radiation patterns.
In the case of antennas with small apertures and operating at large
wavelengths, the transition from near
to far field may occur within the atmosphere: consequently, emission from
the atmosphere in overlapping far field patterns may not be rejected
by the interferometer. These effects have been examined 
by Church (1995) and Lay \& Halverson (2000).
\item (iii)  The short-spacing interferometer may respond to 
nearby environmental emission
in overlapping spillover patterns on the ground. Variations in 
brightness temperature across the ground and because of trees or buildings
would be the cause of a `ground fringe'.
\item (iv) Any interferometer that has a baseline 
component perpendicular to the horizon 
may respond to the discontinuity in brightness 
at the horizon between the sky and ground, if both the 
antennas respond to the interface.
\item (v) Shadowing and overlap between projected apertures may, effectively,
generate a zero-spacing interferometer that responds to the uniform sky,
atmosphere and ground.
\endlist
Some of the above mentioned items are not independent and reflect simply
different viewpoints of the same phenomenon.

Besides these problems, it may also be mentioned here that external
interference is of greater deleterious consequence 
for short-spacing interferometers 
because the associated fringe rate is small.  
Additionally, because the effects of band-width
and time-averaging smearing are smaller for short baselines,  strong
sources (like the Sun) may produce a correlated response even if they
are at large angles from the antenna pointing direction.

Correction for some of the multiplicative effects will vary across the 
primary beam.  In the celestial sphere, the primary beam may be time
varying as a source is tracked across the sky if the effective 
aperture illumination of the shadowed antenna changes as a result of 
changes in the degree of geometric shadowing.
Equivalently, in the spatial frequency (visibility)
domain, the measured visibility is an average over a region of the 
spatial-frequency plane with a time-varying weighting function.  

\section{Observations of the cross talk between ATCA antennas}

The ATCA is an array of Cassegrain-type antennas with shaped reflectors
and forms a Fourier-synthesis radio telescope.  Five antennas, labelled
ca01 through ca05, are movable
along a 3 km  EW railtrack -- that also has a short NS spur -- and may be 
sited at any one of several fixed station locations along the track.  
The main reflectors of the antennas are 22~m in diameter and
the minimum allowed baseline is 30.6~m.  All the cross-talk 
measurements described
below were made between a pair of antennas located on stations that were
30.6~m apart on the EW track.

The Cassegrain antennas have a short focal ratio, with an $f/D$ ratio 
of 0.32, and have axially symmetric optics.  A 2.74-m diameter 
sub-reflector is mounted on a tetrapod above 
the 22-m diameter main reflector.  The reflector optics are `shaped'
to maximize G/T: the ratio of the antenna gain to the 
antenna system temperature; 
therefore, the main and sub-reflector surfaces depart from 
paraboloidal and hyperboloidal.
The sub-reflector is made of solid aluminium; the
main reflector surface is formed by six concentric rings of panels.
Originally, the inner four rings out to 5.865~m radius were made of solid
panels and the outer two rings were made of perforated panels; recently,
as part of a `high-frequency upgrade' of the telescope, the outer rings
have been replaced with solid panels. The panels are
constructed from segments that are bonded to
stretch-formed ribs of `I' section aluminium, rivets connect these
panel sections to the ribs.  The backup structure
supporting the panels is an open truss system.  
The antenna mount is alt-azimuth.

Most measurements described in this section were made  
at 3 \& 6~cm wavelengths. The feed, whose frequency range spans both
these wavelength bands, is a wide-band compact corrugated horn 
at the Cassegrain focus. The feed is followed 
by an ortho-mode transducer which
provides dual linear polarization signals, labelled X and Y, which 
are down-converted 
and used by the correlator to compute the correlation on any 
given baseline.
When the antenna is viewed from the front (face on),
the legs of the tetrapod, which support the sub-reflector, 
are oriented along 0 and $90^{\circ}$ position angle (P.A.)
and the two orthogonal linear feeds may be visualized as picking up the 
X and Y polarization electric field components at 45 and $135^{\circ}$ 
P.A. respectively.
The phase difference between the two linear polarization signal paths
in any antenna is measured by its on-line injected-noise
calibration system and the four cross-correlation
products X$_1$X$^*_2$, Y$_1$Y$^*_2$, X$_1$Y$^*_2$ and Y$_1$X$^*_2$ 
(denoted hereafter by XX, YY, XY and YX respectively) 
between any antenna pair may be converted to Stokes parameter correlations. 
The measurements were made using 128-MHz bandwidths centred at 8640
MHz in the 3-cm band and at 4800~MHz in the 6-cm band.  
The ATCA correlator measures correlation coefficients
over a range of positive and negative lags and the measurements are then 
Fourier transformed
to provide visibility measurements over a set of frequency channels
covering the observing band.  The data described here were acquired 
in 16 independent 8-MHz channels covering the 128-MHz bands. 

\subsection{The observed visibilities in baselines with shadowing}

The `unresolved' continuum source B1741$-$038 was observed
over the HA range $+1$ to $+5$~hr till before the source set 
in the West; during this time the source elevation decreased 
from 60 to $15^{\circ}$.   Visibility data were acquired
using three ATCA antennas -- ca01, ca02 and ca03 -- configured on the
EW track so that ca01 and ca02 were separated by 30.6~m and ca03
was located 400~m down the track to the West.  The array configuration
is depicted in Fig.~1.  Before and as the source set, antenna
ca02 progressively shadowed ca01; ca03 did not at any time block the
aperture of ca02 and, consequently, ca02 and ca03 remained 
unblocked throughout.

\beginfigure{1}
\centerline{\psfig{file=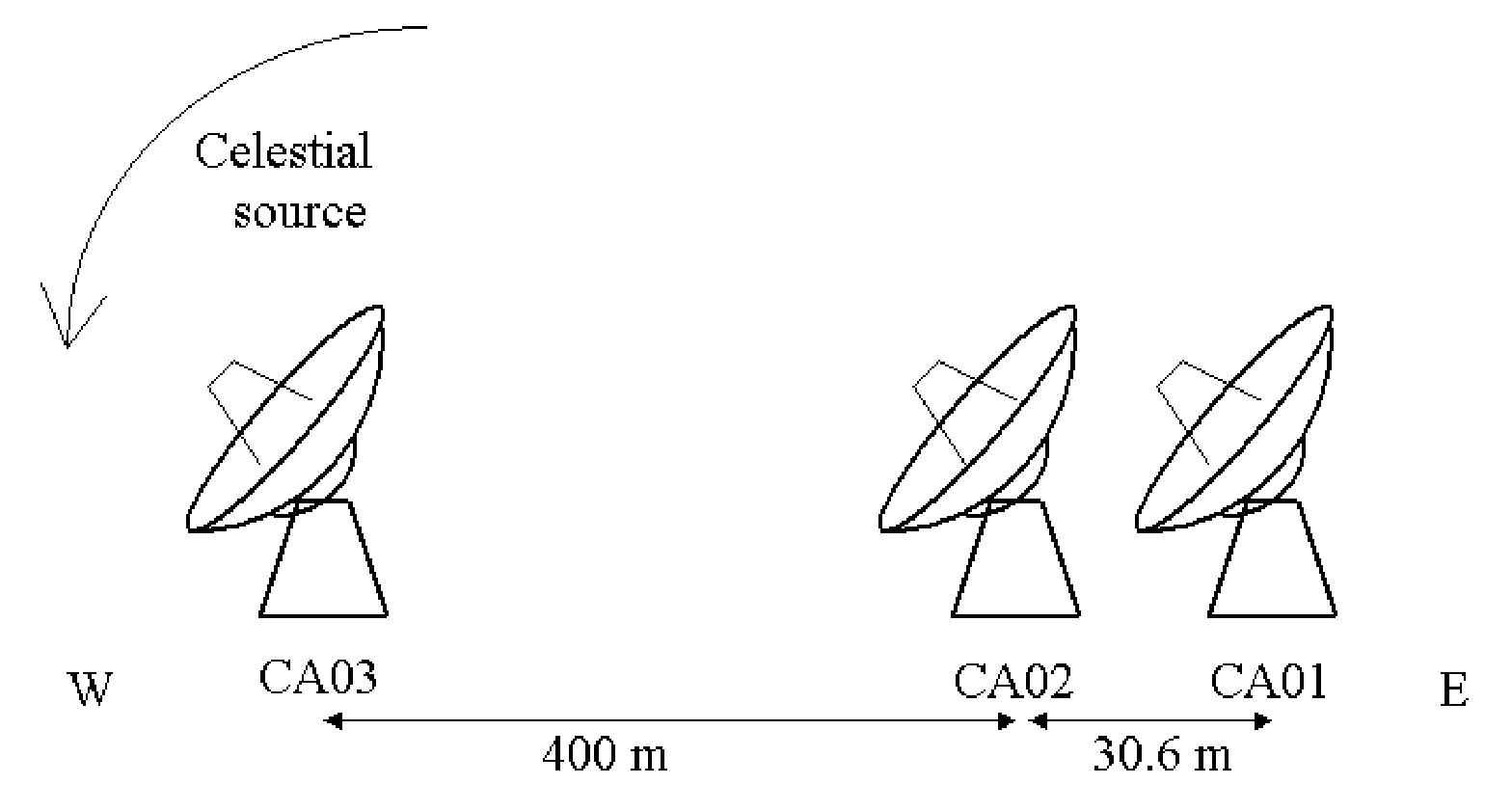,clip=,height=4.5cm,angle=0}}
\caption{{\bf Figure~1.} The array configuration for the observations 
described in Section~3. }
\endfigure

The observed visibility amplitudes at 4800 MHz, calibrated
in flux density and expressed in units of Jansky (Jy),  are shown in Fig.~2.  
It may be noted here that the visibilities have been corrected for 
changes in the system temperatures of the antennas (related mostly to
changes in antenna elevation and shadowing). The panels
show the Stokes I, Q, U and V parameter amplitudes on the baselines 
ca03-ca01 and ca02-ca01 versus the length of the projected baseline
ca02-ca01.  When the projected antenna separation, on the ca02-ca01
baseline, is less than the dish diameter (i.e. 22~m), 
ca02 geometrically shadows ca01. The drop in the visibility amplitude on 
the ca03-ca01 baseline, as antenna ca01 is increasingly
shadowed by ca02, is because the effective collecting area of ca01
is reduced when it is shadowed and the amplitude change may be 
considered to be an `antenna-dependent' effect.
All unshadowed antennas in the array would show a similar drop in 
visibility amplitude in baselines with the shadowed antenna ca01.
The behaviour of the amplitudes on the ca02-ca01 baseline
are observed to be unlike that seen in the ca03-ca01 baseline:  
coincident with the onset of geometric shadowing (when the projected
separation reduces below 22 m), the Stokes I, Q and U
amplitudes on the ca02-ca01 baseline are observed to
diverge from the ca03-ca01 amplitude and the difference grows
as the degree of shadowing increases.  No significant flux density
is observed in Stokes V either before or after the onset of
geometric shadowing.  We believe that the divergence in the Stokes
I, Q and U amplitudes in the short baseline interferometer ca02-ca01
is because of an additive `baseline-dependent' component that is
observed in the baseline between the shadowing and shadowed antennas; we
hereinafter refer to this spurious correlated signal component as cross talk.

\beginfigure*{2}
\centerline{\psfig{file=fig_2.ps,clip=,height=8.5cm,angle=-90}}
\caption{{\bf Figure~2.} Effects of shadowing on visibilities of
a continuum unresolved source.  The panels show Stokes~I, Q, U and V 
visibility amplitudes at 4800~MHz. The data are plotted as a function of the
projected spacing between ca01 and ca02; geometric
shadowing of ca01 begins when this projected antenna separation falls below
22~m. The amplitudes
on the baseline between the shadowed antenna (ca01) and the shadowing
antenna (ca02) are shown using continuous lines and those 
between the shadowed antenna
(ca01) and a distant antenna (ca03) are shown using dashed lines.  
Note that the flux density scale runs from 2--5 Jy for the panel showing
the Stokes~I amplitudes where as the scale runs over the range 0--0.25 Jy
in the panels showing the other Stokes parameter amplitudes.}
\endfigure

\subsection{Spurious additive visibilities on shadowed baselines}

We have examined the additive cross talk, in isolation,
by observing towards a `blank' sky region with a pair of 
antennas in a situation where one antenna shadows the other.  
Using antennas ca01 and ca02, separated by 30.6~m on the EW track,
a blank-field at $-4^{\circ}$ declination (identical to that of the
source B1741$-$038) was observed
as the field set in the west and the western antenna, ca02, increasingly
shadowed ca01 which was located to its east on the railtrack.  The 
observed visibility amplitudes and phases at 
4800~MHz versus the projected antenna separation are shown in Fig.~3.
Cross talk was observed in the ca02-ca01
baseline when the projected separation was less than the
dish diameter (22~m) and the antenna in front, ca02, geometrically
shadowed the antenna at the back, ca01.
The Stokes~I amplitude of the cross talk progressively
increased as the geometric shadowing increased and the projected
baseline reduced; the Stokes~I cross talk phase is
fairly constant in the shadowed regime.  The linearly polarized Stokes~Q
and U amplitudes are, at 4800~MHz, about 20 per cent of the Stokes~I amplitude;
however, the polarization Stokes parameter amplitudes do not have a
monotonic behaviour with increasing shadowing and their phases vary
with the geometry of the shadowing.  A small Stokes~V component
is also detected at 4800~MHz that is about 5 per cent 
of the Stokes~I amplitude.

\beginfigure*{3}
\centerline{\psfig{file=fig_3.ps,clip=,height=8.50cm,angle=-90}}
\caption{{\bf Figure~3.} Cross talk visibilities -- observed on the
ca02-ca01 baseline -- versus the projected antenna spacing in metres.   
The four panels show the cross talk in the four Stokes parameters,
the amplitudes are plotted as continuous lines and the phases are 
shown using symbols.  Note that the vertical scale runs from 0--1
for Stokes I and runs over the range 0--0.25 in the panels showing 
the other Stokes parameters. The onset of geometric shadowing
occurs at 22~m.}
\endfigure

In an attempt to correct the visibility on the short baseline ca02-ca01
for the additive cross-talk, the complex cross-talk 
visibility observed on the blank field 
were subtracted (vectorially) from the visibilities observed 
on the calibrator source at the same geometric shadowing.  
Following the subtraction of
the additive cross talk, the ca02-ca01 baseline would be expected to
have only antenna-based errors associated with the blockage of the 
aperture of ca01 and these multiplicative errors would be the same as those
afflicting the ca03-ca01 baseline.  
The subtraction was successful within the errors in the measurements:
the corrected ca02-ca01 visibilities in all the four Stokes
parameters were observed to have a behaviour similar
to those on the ca03-ca01 baseline.  

It is remarkable that the complex spurious signal
observed towards the blank field (Fig.~3), in all the four Stokes
parameters, are 
the same as the additive cross talk component present in the
observations of the unresolved source (Fig.~2).
The inference is that the cross talk is not a coupling of the incident
power, from the source(s) at the field centre, via a new path which
is created because of shadowing. The cross talk is a complex additive 
(and not a multiplicative)
error on the short baseline in each of the Stokes parameters,
it has a high degree of linear polarization, and the cross talk
amplitude and phase depend on the shadowing geometry.

\section{Characterization of the cross talk}

The additive cross talk is observed, using the ATCA spectral line 
correlator, to be a continuum signal and not a
narrow-band line.  The cross talk originates as broad-band noise
and is not, for example, a local oscillator or any such tone-like
signal generated in the receiver electronics.

With a pair of antennas in a shadowed configuration, both
antennas pointing towards the same direction and with the
geometric delay compensated for the antenna pointing centre,
we have observed the cross talk with a 128-MHz
bandwidth and examined the lag-spectrum with 3.9~ns steps and over 
a $\pm 125$~ns range.  The cross talk appears at `zero' lag, i.e., with 
a delay appropriate for a continuum source at 
the antenna-pointing centre.  From the lack of any detectable
phase gradient across the 128 MHz band, we infer that
the cross talk appears to be coupling
into the signal paths of the adjacent antennas with a delay that is the
same as the geometric delay within $\pm 0.3$~ns 
(corresponding to a free-space propagation path difference 
of less than 10 cm).
We have examined the cross-talk phase in widely separated observing bands
and observed that the phase has a 
complex and slow variation with frequency.  The total phase change
is about $2 \pi$ radians over the range 1384--9152~MHz. The observed
phase variation with frequency is consistent with a small 
difference between the geometric delay and the relative delays 
between the cross talk signal paths to the two antennas; 
the observations imply that this delay path difference is at most several cm.
Operating the receivers in a spectral-line mode and with a narrow
bandwidth, we have examined the correlation between the antenna signals
over a range $\pm 128~\mu$s about the geometric delay
and do not see any other signal over this large range.

The amplitude of the cross talk signal decreases with increasing frequency:
in the range 1384--8640~MHz, the amplitude, calibrated in Jy, roughly
decreases as $S_{\nu} \propto \nu^{-1}$, where $\nu$ is the observing
frequency.  The cross talk has an equivalent 
antenna temperature which also decreases as $T_{a} \propto \nu^{-1}$.
At cm wavelengths, the cross talk is observed at a level
corresponding to a correlation coefficient
of about 0.1 per cent and the amplitude corresponds to a geometric mean 
antenna temperature of about 50~mK.

The observed cross talk amplitude and phase 
remain unchanged when the noise diodes, which inject
noise for calibrating the receiver gain and system temperature, are
switched off; this is a confirmation that the calibration noise 
diodes are not the source of the power for the `unwanted' correlation.

The cross-talk signal appears as if it originates in the sky from 
the antenna-pointing 
direction and with a delay appropriate for signals from that sky direction.  
Therefore, we believe that the cross talk is not due to
the direct coupling of common ground radiation into the two receivers.  
Additionally, we do not believe that the spurious correlation is 
because of any leakage of receiver noise from the front antenna 
and into the signal path of the antenna behind.  

The cross talk is because of a common mode signal that enters the
two receivers located in the adjacent antennas.  The sharpness of the
response in the lag domain argues in favour of a dominant and perhaps a 
single coupling path and against a scenario where the coupling  is 
via multiple scattering paths, unless these multiple paths have the
same differential delays.  Coupling because of scattering off, for
example, the sub-reflector support structures may be expected to be spread
over a range of delays.     

\subsection{Dependence of the cross-talk amplitude on the shadowing 
configuration}

As the projected baseline between a pair of ATCA antennas 
decreases to values below the antenna
diameter, cross talk is observed and the Stokes~I amplitude of the cross
talk increases as the projected baseline reduces further.
With two antennas on stations separated by the minimum spacing of 30.6~m
on the EW track, geometric shadowing would be a maximum (and the
projected baseline would be a minimum) when the antennas
are pointed along the track towards azimuth of 90 or 270$^{\circ}$ and 
the antenna elevations are at their lowest; the 
cross-talk amplitude in Stokes~I is observed to systematically decrease
away from these sky positions.  We observed 
the cross-talk amplitudes when the 30.6-m baseline 
interferometer was pointed towards a mosaic of azimuth-elevation positions
in the range azimuth: 220--320$^{\circ}$ and 
elevation: 15--55$^{\circ}$.
The variation in the Stokes~I cross-talk amplitude, as well as in the 
XX and YY polarization products, versus offsets in the pointing
from azimuth 270$^{\circ}$ and elevation 0$^{\circ}$, is shown in Fig.~4.

\beginfigure{4}
\centerline{\psfig{file=fig_4.ps,clip=,width=8.0cm,angle=00}}
\caption{{\bf Figure~4.} Contours of the cross talk amplitude at 4800~MHz:
the upper panel shows the Stokes~I amplitude, the lower left panel
is for the XX polarization and the lower right panel is of the
YY polarization amplitude.
The axes represent the offset in pointing direction from $0^{\circ}$
elevation and $270^{\circ}$ azimuth.  The maximum cross-talk amplitude was 
observed at an azimuth of $270^{\circ}$ when the antennas were 
at their lower elevation limit of $15^{\circ}$;
contours are shown at intervals of 10 per cent of this maximum amplitude.  
Geometric shadowing occurs for offsets within the dashed circle
and no significant cross-talk amplitude is observed when the offsets
cause the pointing to lie outside this circle.}
\endfigure

When the projected baseline exceeds the antenna diameter and
there is no geometric shadowing, the cross talk amplitude 
in Stokes~I as well as in the XX and YY polarizations
drop to zero (or below our detectablity).

The Stokes~I cross-talk amplitude across the ranges of azimuth and elevation
offsets is observed to have a circular symmetry about the direction of 
complete shadowing: it is thus dependent only on 
the length of the projected shadowed baseline
and independent of the position angle of the projected baseline.
The symmetry adds weight to the argument against 
ground radiation as a source of the cross talk.  

The XX and YY polarization products show different 
azimuth-elevation dependences 
that depart from such circular symmetry: the cross talk is linearly polarized
with XX cross-talk amplitude exceeding the YY amplitude for positive
azimuthal offsets and vice-versa.  This implies that when a pair of
antennas, which are in a shadowed configuration, are viewed face on 
and from along their pointing direction, the net 
cross talk at 4800~MHz is partially linearly polarized and has an
equivalent net E-field that appears as if it is 
oriented perpendicular to the line joining their centres.

\beginfigure{5}
\centerline{\psfig{file=fig_5.ps,clip=,width=8.0cm,angle=-90}}
\caption{{\bf Figure~5.} 
As the projected baseline between two
antennas that are in a shadowed configuration decreases, the
observed cross-talk amplitude (shown as a continuous line joining the
box symbols) is compared with the area of the back antenna
that is geometrically shadowed (dotted line) and the length of the
rim of the front antenna that appears projected on the back antenna
aperture (dashed line).  The observed cross-talk amplitudes displayed
are at 4800~MHz; the vertical scales for the rim length and shadowing
area are arbitrary.}
\endfigure

The variation in the cross talk amplitude on a 30.6-m baseline is shown
in Fig.~5 versus the length of the projected baseline.  These data were
obtained at 4800~MHz.  As noted earlier, the cross talk signal is significant
only when the antennas geometrically shadow and the projected baseline
length drops below 22~m.  We have plotted, in the same figure, the 
area of geometric shadowing (the area of the overlap segment) versus
the projected baseline using a dotted line.  Also plotted (using 
a dashed line) is the length
of the rim of the front antenna, which appears across the aperture of the 
back antenna, when the antennas in a shadowed configuration are viewed
face on and from along their pointing direction.  The comparison indicates
that the cross talk arises owing to a coupling associated with the
area of geometric shadowing; however, the effect is not uniform and,
at least at 4800~MHz, is biased towards the outer edge of 
the overlap segment.

\section{The distribution of the cross talk across the apertures}

The cross talk is because of an `unwanted' signal
that enters the receivers of the antennas, which are in a configuration where
one of them geometrically shadows the other.  The interferometer response
is because of coherence between these signals at the two antennas: it 
may be noted here that the interferometer response does not require a
spatial coherence within the individual apertures, it only requires that
there be coherence between the fields incident on the two apertures.
The radiation field that causes the cross talk would have a distribution
across the apertures of the two antennas and examining these
distributions is important to understanding the source of the cross
talk and the nature of its coupling into the two signal paths.
When an incident electromagnetic wave is spatially coherent across
the space containing a pair of antennas, the spatial distribution of the field
which couples into the individual antennas may be measured 
using standard holographic methods (Bennett et al. 1976; Scott \& Ryle 1977).  
The cross talk we observe at the ATCA in shadowed configurations 
may not be due to a wavefront having transverse spatial coherence.
However, it should be still possible to seek the equivalent 
aperture plane distributions of the
cross-talk signal through holographic reconstruction 
if the signal couples into the aperture planes of the antennas in a 
distributed manner.

\subsection{ 1-dimensional analysis of the distribution in the cross talk}

\beginfigure*{6}
\centerline{\psfig{file=fig_6.ps,clip=,height=9.0cm,angle=-90}}
\caption{{\bf Figure 6.} 
Two antennas 30.6-m apart E-W point 
nominally towards AZ = $270^{\circ}$
and EL = $20^{\circ}$.  The variation in amplitude (continuous 
line) and phase (symbols) of the cross
talk at 4800~MHz are shown as the azimuth of one antenna is offset from
the nominal direction keeping the antenna elevation fixed.  
The panels on the left are for the case where the antenna in front scans in
azimuth; those on the right correspond to the case where the antenna at
the back scans.  The three panels on either side separately show the
Stokes~I visibilities and the XX and YY polarization products.
}
\endfigure
\beginfigure*{7}
\centerline{\psfig{file=fig_7.ps,clip=,height=9.0cm,angle=-90}}
\caption{{\bf Figure 7.} 
As in Fig.~6, with  offsets now applied in elevation instead of azimuth.
Both antennas point towards a constant azimuth of $270^{\circ}$ and
the offsets are made in elevation about the nominal EL = $20^{\circ}$.
}
\endfigure

A pair of antennas were configured 30.6~m
apart on the E-W track and pointed along the track at azimuth $270^{\circ}$
and elevation $20^{\circ}$.  Geometric path delays and phases were
compensated for this sky direction.  In this shadowed configuration, 
a cross-talk signal is observed.  Keeping one of the two antennas 
fixed on the above pointing, the other antenna pointing was varied to scan 
across a range of offsets
in azimuth and elevation around the nominal pointing direction and
the complex cross-talk visibility were recorded.
In Figs.~6 \& 7 we show the variation in the cross talk amplitude and phase 
at 4800~MHz as a function of the offset from the nominal pointing
direction (azimuth = $270^{\circ}$; elevation = $20^{\circ}$);
Fig.~6 shows the visibility variations for pointing 
offsets in azimuth and Fig.~7 displays these for 
elevation offsets from the nominal position.  In each of these figures,
the panels on the left represent the case where the pointing of the 
antenna at the back was kept fixed at the nominal position while the 
antenna in front executed the scan patterns, the panels on the right
are for the case where the antenna in front had its pointing fixed towards
the nominal direction while the antenna behind executed the scans
in azimuth and elevation.  For each of these cases, 
the measured Stokes~I visibilities as well as the XX and YY polarization
products are shown in separate panels.
The visibility phases shown in the figures represent the phase of the
scanning antenna with respect to that of the stationary antenna (pointing
towards the nominal direction); this differential phase 
has had the phase corresponding to the geometric 
path delay for the nominal pointing direction subtracted.

First, the cross-talk amplitude falls off fairly rapidly
when a relative pointing offset is introduced between the
shadowing and shadowed antennas. 
We have derived first-order estimates of the effective 
aperture size contributing to the cross talk
from the apparent beamwidths of the scan patterns in azimuth and elevation.
These estimates suggest that 
almost the full horizontal extent and only about
half of the vertical extent of the physical apertures may be relevant.
This in turn suggests that the
observed cross talk is a vector sum of a large number of coupling
paths and that introducing relative pointing offsets between the antennas
result in phase distributions across these contributions causing
cancellations in the vector summation.  The coupling is via multiple
paths that are distributed over the aperture; however, 
the arguments presented earlier require these paths
to be very close in their differential delays.

The Stokes~I azimuth scan patterns are remarkably symmetric and the 
same (except for the phase sign change) 
irrespective of whether the antenna in front or that behind is offset
from the nominal pointing.  The XX and YY polarization products,
for this case where the offsets are made in azimuth,
are not only asymmetric but are also shifted in that the peak response 
occurs at positions offset from the nominal pointing direction. 
However, they have the symmetry property in that the cross talk measured
in the XX polarization product for a positive azimuth offset is the
same as the cross talk measured in the YY product for a negative offset.
Comparing the scan patterns observed in Stokes~I, XX and YY 
for the two cases where the front antenna scans and where the antenna
behind scans, the cross-talk amplitude and phase 
observed for positive offsets to the front
antenna are the same as the cross talk observed for negative
offsets to the antenna at the back.  
In other words, when each pair of scans for a given 
product (Stokes~I, XX or YY)
is viewed as a function of the relative azimuth offset 
with respect to the pointing of
a fixed reference dish, say, of the front antenna, the 
scan amplitude profiles are same
irrespective of which dish-pointing was offset, while 
the phase profiles change sign.  
And the profiles in XX are same as for their YY counterparts, 
except for the flip of the azimuth-offset axis.

The elevation patterns (Fig.~7) do not share many of the
symmetry properties of the azimuth scans.  First, we observe only
marginal differences between
the patterns in Stokes~I and in the XX and YY polarization products.
All the patterns are asymmetric.  Additionally,
the cross-talk amplitude, in Stokes~I as well as in 
the XX and YY polarizations,
peak at a pointing offset from the nominal; i.e., the cross talk
peaks for an antenna configuration in which the front and
back antennas are slightly offset in their relative elevation pointings.
The elevation scan patterns share a symmetry property with the
azimuth scans in that the cross-talk amplitude for
positive elevation offsets applied to the front antenna is the same
as that for negative offsets to the back antenna. 
Stated differently, the scan amplitudes (for products Stokes I, XX or YY)
are about same, irrespective of which dish-pointing was offset, 
and only depend on the differential pointing, say, of the back antenna
with respect to the front antenna.
The cross-talk phase patterns in these two cases, however, have 
very different slopes (versus offset)
depending on whether the offsets are made to the antenna at the front
or behind.

To summarize the symmetry property common 
to the azimuth and elevation scans:
considering azimuth offsets, the cross talk is the same irrespective
of whether the antenna at the front is offset to the right or the antenna
at the back is offset to the left; for elevation offsets, tipping the
antenna in front backwards has the same effect on the cross-talk amplitude
as tipping the antenna behind forwards.  These symmetries are 
despite the fact that the antenna at the back is shadowed
and only a part of its aperture would be illuminated by radiation incident 
from the pointing direction.  The implication is that the cross-talk
signal couples from one antenna to the other, so that the cross talk
depends only on the relative orientations of the pair of antennas:
the cross talk is not a coupling of radiation from an external
source independently into the apertures of the two antennas.  The
common-mode signal likely originates in the antenna in front, couples
as a constant signal into the receiver of the front antenna, and
propagates across space into the receiver of the antenna behind
via the reflectors of that antenna.

It may be noted here that we adopt the convention that the visibility
phase $\phi_{ab}$ of any baseline vector from antenna $a$ to 
antenna $b$ is the phase corresponding to the differential delay: the
delay experienced by the signal arriving at antenna $b$ minus the delay 
experienced by the signal arriving at antenna $a$.  Therefore, the negative
slope in the phase of the antenna in front w.r.t. the antenna at the back,
as the antenna in front is moved up in elevation (panels on the left 
in Fig. 7), and the positive slope in the phase of the antenna at 
the back (panels on the right in Fig. 7), imply that the cross-talk 
signal couples between the apertures of the two antennas via the
upper parts of the antenna in front and the lower parts of the antenna
at the back.  

\subsection{ A 2-dimensional analysis of the distribution in the cross talk}

\beginfigure*{8}
\centerline{
(a)\psfig{file=fig_8a.ps,clip=,height=8.0cm,angle=-90}
(b)\psfig{file=fig_8b.ps,clip=,height=8.0cm,angle=-90}
}
\caption{{\bf Figure~8.} 
The 2D Fourier transformation of the distribution in cross talk 
over antenna pointing. The measurements are in Stokes~I and 
at 8640~MHz.  In the case of the panel on the top, the
visibilities input to the transform were acquired while the front antenna
executed a 2D raster in azimuth \& elevation while the antenna at the
back pointed towards the nominal AZ = $270^{\circ}$, 
EL = $20^{\circ}$ direction.  The panel at the bottom corresponds
to the case when the back antenna executed the raster scans.
The panels show the distribution in the transform amplitude and the 
plot coordinates are in units of metres.  On the top panel, a grid 
showing the layout of the panels of the antenna in front is overlaid, 
the outer and inner limits of the panels of the antenna at the back 
are also shown; these overlays are projections perpendicular to the 
nominal pointing direction. On the bottom panel, the layout of the
back antenna panels and the outline of the front antenna aperture
are overlaid. Contours are at 15, 30, 45, 60, 
75 and 90 per cent of the peak; 
the r.m.s. noise is at a level of 3 per cent of the peak.
}
\endfigure

To investigate further, the offsets to the front and back 
antennas were made to cover a 2-dimensional (2D) grid of azimuth 
and elevation positions: elevation offsets
covered the range $\pm 1.3$ degrees, azimuth offsets covered the range
$\pm 1.235$ degrees.  These offsets were made, to one of the
antennas at a time, while the other had its pointing fixed towards the
nominal AZ = $270^{\circ}$, EL = $20^{\circ}$ direction.
Complex visibilities, calibrated in Jy, were 
recorded at each pointing offset.  The visibilities were recorded,
simultaneously, at 4800 and 8640 MHz and in all the four polarization 
products, XX, YY, XY and YX. It may be noted here that the calibration
in amplitude and phase were made using an unresolved source and towards
a sky direction where there was no geometric shadowing.  The 2D grid of 
complex visibilities, which were a function of relative sky-angles (radians),
were then Fourier transformed to produce a complex grid of field
distribution estimates, which are distributed in the conjugate 
variable that is spatial coordinates with units of wavelengths.  
The spatial coordinates following the transforms 
at each frequency were converted to units of metres using the appropriate
wavelength. 

In Fig.~8 we show the spatial distribution of the field amplitudes 
that were obtained by a Fourier transformation of the 
Stokes~I visibilities recorded 
at 8640~MHz; we display the field distributions computed 
via transformations of the complex visibilites recorded while
the front antenna executed a 2D raster in azimuth-elevation
coordinates and, separately, while the antenna at the back
executed the same scan pattern.  Before the transformation to 
spatial coordinates, the
visibilities were tapered using a Hamming window.  Consequently,
the field distribution in spatial coordinates have a point spread
function (PSF) with FWHM 1.4~m at 8640~MHz;
the sidelobes of the PSF are less than 1 per cent of the peak. 

There is considerable fine structure in the correlation 
amplitude distributions.
It may be noted here that most of the structure observed in the 
amplitude distributions shown in the two panels of Fig.~8 are 
genuine and have a high significance statistically: the lowest 
contours plotted are at five times the standard deviation of the image noise.
The distribution obtained from the visibilities recorded while the
front antenna scanned has a striking similarity, even in details, 
to the distribution for the case
where the back antenna executed the raster scan.  Both these
images were constructed from independent visibilities, observed
at different times, but between the same pair of antennas and with the
same antenna in front at both times.  The similarity in the details
of the distributions is additional confirmation that the
observed structure is well above the image noise.  

The Stokes~I distribution does
not show any obvious relationship to the panel structure, but does show a 
clear correspondence with the concentric annular structure of the front dish.
For example, there
are nulls in the amplitude distribution at specific radii measured
from the vertex (centre) of the aperture of the front antenna. 
In the 8640-MHz data, the nulls are
located at radial distances of 5.5 and 8.7~m and at these radii the
phase flips by about 180 degrees.  At 4800~MHz, a phase flip is
observed at a radius of 4.6~m.  It may be noted here
that the detailed structure in the distribution has been observed
to differ when the measurements are made using a different pair 
of antennas.  The overall distribution is observed to vary also with frequency.

The holographic imaging of the cross talk confirms and extends 
many of the inferences derived from the azimuth and elevation scan data.
First, it may be infered that 
the cross talk does not arise due to a coherent wavefront incident on 
the antenna pair from the far field and parallel to the pointing direction.
If this were the case, the 2D Fourier Transforms of the cross talk
visibilities that were recorded while the two antennas separately 
executed azimuth/elevation scans would have
given us images of the entire aperture for the case where
the front antenna scanned and images of the unblocked aperture for the
case where the shadowed antenna scanned.  Additionally, the cross talk is
not a consequence of the scattering by the front antenna, of 
power from a plane wavefront that is incident on the front antenna, to
the second antenna behind.  

The Fourier transforms of the 2D azimuth/elevation raster scans, made
separately for the antenna in front and the antenna behind (see 
Fig.~8), are the same.  This is consistent with the interpretation
in which the common-mode signal enters one of the antennas -- the antenna
in front -- as a constant
coupling independent of the small variations in the azimuth/elevation
pointing of that antenna during the raster scan.  This common-mode signal
is radiated by elements of that antenna and couple into the aperture
of the antenna behind.   The distribution of the
field sharply cuts off at the edges of the overlap region.  The 
cross-talk field is also zero in the central regions 
of the two apertures that are
occupied by the feed housing at the vertices of the main reflectors.
The holographic imaging indicates that the cross-talk coupling is
localized to the segments of the apertures that would appear to overlap
as viewed from the direction towards which the antennas point;
the coupling is confined to the areas of the apertures that geometrically 
shadow.  Additionally, it rules out any transverse coherence in the coupling
field on scales more than a small fraction of the aperture
(i.e., more than a few Fresnel scales).   The similarity in the
distributions derived from data in which the front antenna scanned
and that in which the back antenna scanned indicates a one-to-one
correspondence between points on the front and back antenna apertures;
the common-mode field might be viewed as propagating from the front
antenna aperture to the back antenna aperture along lines parallel
to their optical axes.  Additional evidence for such a propagation path 
is the observation that the signal enters
the receivers with a relative delay that is the same as that for a
celestial source in the direction that the antennas point.
A coupling via propagation between the overlap portions of the
two apertures is consistent with the observations that the variations 
in cross-talk visibility only depend on the relative pointing
between the two antennas and that the two Fourier transforms in Fig.~8
display the same distribution.

The images in Fig.~8 show that the signal arrives at the back antenna
exclusively from the region of the back surface of the front antenna that
shadows the antenna behind.  In other words, the common-mode signal
propagates to the aperture of the back antenna from the entire 
back surface of the
front antenna. Given that the Fresnel scale is small (less than 2~m),
the power associated with the cross talk can be considered as 
propagating more or less along lines parallel to the
optical axis of the back antenna and will reach the receiver of 
that antenna after reflections off the main and sub-reflectors
of that Cassegrain antenna.  As viewed by the 
antenna behind, the source of the common-mode signal is in the 
main reflector surface of the antenna in front.

If the source is mainly spatially incoherent emission 
distributed across the surface
of the main reflector of the antenna in front, the observation that the
cross talk appears at zero lag (after the geometric path delay corresponding
to the pointing direction is removed) once again leads to the
inference that the emission must couple
across the main reflector surface in a distributed manner.    

\subsection{ The distribution in the polarization of the cross talk}

\beginfigure*{9}
\centerline{
(a)\psfig{file=fig_9a.ps,clip=,height=8.0cm,angle=-90}
(b)\psfig{file=fig_9b.ps,clip=,height=8.0cm,angle=-90}
}
\caption{{\bf Figure~9.} 
The 2D Fourier transformations of the XX (top panel) and YY (bottom
panel) polarization product measurements of the cross talk
at 4800~MHz.  In both cases, the
visibilities input to the transform were acquired while the front antenna
executed a 2D raster in azimuth and elevation while the antenna at the
back pointed towards the nominal AZ = $270^{\circ}$, 
EL = $20^{\circ}$ direction.  
The panels show the distribution in the transform amplitude and the 
plot coordinates are in units of metres.  A grid 
showing the layout of the panels of the antenna in front is overlaid, 
the outer and inner limits of the panels of the antenna at the back 
are also shown.  Contours are at 10, 20, 30, 40, 50, 60, 70, 80 
and 90 per cent of the peak.
}
\endfigure
\beginfigure*{10}
\centerline{
(a)\psfig{file=fig_10a.ps,clip=,height=8.0cm,angle=-90}
(b)\psfig{file=fig_10b.ps,clip=,height=8.0cm,angle=-90}
}
\caption{{\bf Figure~10.} 
The 2D Fourier transformations of the XX (top panel) and YY (bottom
panel) polarization product measurements of the cross talk
at 8640~MHz.  The coordinates, contour intervals and labelling
are as in Fig.~9.
}
\endfigure

The symmetry properties of the scan data, in
Stokes~I as well as in the polarization products,
and the images of the distribution in the field amplitude across the aperture 
suggest that the cross-talk coupling between the two antennas is 
described via a single coupling distribution: the measurements made
by scanning the antenna in front or the antenna at the back are both
a measurement of the same coupling field. To investigate the polarization
in this coupling, we have examined the distributions in the aperture field
of the XX and YY polarization products.  
In Figs.~9 and 10 we show these distributions, at 4800 and 8640~MHz
respectively,
for the case where the front antenna executed the raster scan.  The 
complex visibility data that were transformed to produce these images were
not tapered and, therefore, in the case of the distribution 
at 8640~MHz displayed in Fig.~10, the spatial resolution is about a 
factor of two better as compared to the corresponding Stokes-I 
distribution shown in Fig.~9; however, the PSF now has higher sidelobes.
The polarization images at 8640~MHz (Fig.~10) have a PSF resolution 
that is about a factor of two finer than
that for the polarization images at 4800~MHz (Fig.~9).
These polarization product images represent the cross-talk
field distribution projected on the antenna apertures and the view
is of the antenna apertures face on.  
In this face-on view, the XX linear polarization corresponding to 
the feed is oriented at a position angle of $-45^{\circ}$ and the
YY polarization is at $+45^{\circ}$ with respect to the vertical.

Both of the XX and YY polarization images have distributions that
cover the entire overlap segment. 
There is a wealth of detailed structure in the field distribution
across the aperture.  
The most striking aspect is that the structures
appear elongated: (i) the elongations in each of the images are 
orthogonal to the orientation of the feed polarization, and (ii) the
elongated structures show a remarkable correspondence with the
panel edges.   
At both frequencies, the XX-polarization images show enhanced
brightness along the radial inter-panel gaps on the left halves
of the overlap sectors whereas the YY-polarization images show
enhanced brightness along the radial inter-panel gaps on the
right halves of the overlap sectors.
At 8640~MHz, the XX-polarization image shows enhanced emission
along the circumferential inter-panel gaps in the right half
of the overlap segment and the YY-polarization images shows
enhanced emission along the circumferential inter-panel gaps
in the left half of the overlap segment.
At 4800~MHz, the emission in the right half of the XX-polarization 
image and the left half of the YY-polarization image 
are diminished at the inter-panel gaps and appear to be peaked
at the panel centres.
It may be noted here that this correspondence is 
with the layout of the front antenna panels, not those of the
antenna behind, consistent with our earlier inference based on 
the Stokes-I images.  

A slot between the plates that form
a reflector surface may couple radiation across the surface, or act as
a slot antenna; the resulting emission will be linearly
polarized with an orientation perpendicular to the slot.  
Such radiation, with polarization perpendicular to panel edges
(i.e., panel gaps) and that perpendicular to the slots (within the panels),
will be picked up by polarization products that are oriented perpendicular
to the panel edges and slots.  This is what is observed in Figs.~10 and 11. 
The polarization images suggest that the 
polarization products respond to cross talk associated with 
those panel edges/gaps/slots that are oriented 
orthogonal to the feed polarization
orientation.   In the images made at 8640~MHz (Fig.~10), 
the XX-polarization product (corresponding 
to $-45^{\circ}$ P.A.) appears to respond to radiation from the
circumferential edges of the panels in the right half of the overlap
segment and radial edges of the panels in the left half of the 
overlap segment.  The YY polarization product, which is oriented
orthogonal to the XX product and at a P.A of $+45^{\circ}$, 
picks up radiation from the other sets of edges/slots.  
The dominant source of the cross-talk radiation, at 8640~MHz, 
appears to be linearly polarized emission that couples into the 
receivers from the inter-panel gaps distributed over the part of the
front antenna main reflector that geometrically overlaps the
antenna behind.  

It may be noted here that there appears to be
significant emission, in the
right half of the XX-polarization and left half of the YY-polarization
images at 8640~MHz, that are off the circumferential inter-panel gaps.
Additionally, at 4800~MHz, the emission in these regions appears
centred on the panels.  These imply that a sigificant cross-talk 
component, with a linear polarization oriented circumferentially, does
originate away from the inter-panel gaps and from the panels themselves.

The main reflector panels are bolted to the backup structure and there 
are gaps of about 1--2~mm between the solid panels.  In order to examine
whether the cross talk is purely owing to a 
leakage of radiation through these
gaps, we used Aluminium tape (3M~425 Aluminium foil tape) 
to seal all the inter-panel gaps over one quarter
of the main reflector of the antenna in front; 
with the front antenna viewed face on, all inter-panel gaps 
in the the entire upper left quadrant were covered.
The Al tape was stuck on over the paint; however, the tape 
would be expected to provide electrical
continuity at the observing frequency.  Raster scans were executed with
the Al tape stuck on.  Fig.~11 shows the results for this case: the
aperture field distributions obtained as Fourier-transform 
images of the XX and YY 
polarization products at 8640~MHz.  It may be noted here that the antenna pair 
used for this measurement was not the same as the pair used for producing 
Figs.~8 or 10.  

\beginfigure*{11}
\centerline{
(a)\psfig{file=fig_11a.ps,clip=,height=8.0cm,angle=-90}
(b)\psfig{file=fig_11b.ps,clip=,height=8.0cm,angle=-90}
}
\caption{{\bf Figure 11.} 
Same as Fig.~10 except that in this case the inter-panel gaps in 
a quarter of the front antenna were sealed with Aluminium foil tape.
In this face-on view of the antenna apertures, the tape covered the
gaps in the top-left quarter of the front antenna.
}
\endfigure

The most striking effect of the Al tape is that the cross talk
aperture field, in the quadrant where the tape was stuck on, is
not detectable in the XX-polarization product.  The YY-polarization 
product detects significant emission from the aperture
region where the inter-panel gaps were sealed; however, the image
of the field in this region is different as compared to the
case without the tape in that the distribution does not follow
the circumferential gaps between the panel rings.  Instead, with the
tape sealing the gaps, the cross talk field appears to avoid the
panel boundaries.  The experiment with the Al tape confirms
the inference that the coupling of
radiation across the main reflector surface is not wholly due to
the inter-panel gaps.  There is an additional component distributed
over the panel surfaces; this surface component is linearly polarized
with an E-field oriented in the radial direction.  

The ATCA panels are solid surfaces; however, as mentioned in Section~3, 
they are not constructed from single
metal sheets.  The panels are made of circumferential segments that
are riveted onto I-section backup ribs. Before assembly, a layer
of epoxy resin was applied to the rib so that the panels could take on the
shape of the assembly mould -- a bed of bolts -- without stresses
on the stretch-formed ribs. The epoxy serves as a gap filler and
this manufacturing process allows gaps between the panel surfaces
and the backup ribs.  We observe that the surface component has
a polarization perpendicular to the slits between the sheets forming
the panels just as the edge component has a linear polarization
perpendicular to the gaps between the panels.  It appears that the 
intra-panel slits between the panel sheets, in addition to the 
inter-panel gaps between panels, are both associated with the
cross talk.  

At 8640~MHz, the distributions in Fig.~10 show that overall the XX amplitude
is greater in the right half of the segment where as the YY amplitudes
are greater on the left side.   We observe that in the lower frequency
observations, at 4800~MHz, the XX amplitudes are greater in the left
side and the YY amplitudes are greater on the right side, which is
the opposite of what is observed at 8640~MHz.
At 4800~MHz, the intra-panel component, with a polarization oriented radially,
dominates the cross talk in the right-half segment of the XX polarization
and the left-half segment of the YY polarization. In these
regions, this intra-panel contribution dominates that arising
from the circumferential inter-panel gaps.  At the higher frequency
of 8640~MHz, the intra-panel contribution is sub-dominant. 
These comparisons
between the images in Figs.~9 \& 10 indicate that the relative
dominance of the intra- and inter-panel contributions, 
the relative strengths of the cross-talk signal arising from the 
radial and circumferential inter-panel gaps, and also the
radial variation in their individual strengths, are significantly 
dependent on wavelength, at least at cm wavelengths.

\subsection{ The phase associated with the components of the cross talk}

The histograms in Fig.~12 show the distribution in the 
observed phases in the case where the upper left quadrant (face-on view)
of the front antenna had Al tape covering the inter-panel gaps.
The phases in the right half of the YY-polarization image,
shown using a dashed line, correspond to the phase of the 
cross talk arising from the radial inter-panel gaps and 
show a peak at about $+150^{\circ}$.  The phases in the left half of the
YY-polarization image, shown using a dotted line, correspond to
the phase of the cross talk arising from the intra-panel slits
and peak at about $-70^{\circ}$.  The histogram of the phases in the
right half of the XX-polarization image, shown as a continuous line,
corresponds to phases of the cross-talk signal from intra-panel slits
and circumferential inter-panel gaps; this histogram is seen 
to be a combination of the dotted and dashed histograms indicating
that here we have a bi-modal distribution and that the cross talk
arising from the radial and circumferential inter-panel gaps have 
similar phases.  The histogram distributions imply that there are
different phases associated with the cross-talk arising from
the inter- and intra-panel slots and that they are both very different from
that expected for a point source at the interferometer phase centre. 
The intra-panel slits have I-section backup members running along the slits on
the backsides of the panels, the inter-panel gaps also have structural
members running along the gaps; therefore, there is no direct line-of-sight
that is parallel to the optical axis and from the slits or gaps to the 
aperture plane of the antenna behind.  These structural backup members
may be responsible (at least in part) for the observed phases associated
with the contributions from the gaps and slits.  It may also
be noted here that the observed phases for the different cross-talk
contributions vary with the observing frequency.  The observed
phase differences and the frequency dependance may be related to
the path delays owing to scattering off the backup structure; as discussed
in Section~4, such a path delay of a few cm is expected from the
observed variation in the phase of the cross talk over a decade in frequency. 

\beginfigure{12}
\centerline{\psfig{file=fig_12.ps,clip=,height=6.0cm,angle=-90}}
\caption{{\bf Figure 12.} 
Histograms of the observed phase distributions -- at 
8640~MHz -- across the aperture
in the case where the inter-panel gaps in a quadrant of the front 
antenna aperture was covered with Al tape.  
The continuous line shows the histogram of the XX-polarization
phases in the right half of the face-on aperture (the top panel
in Fig.~11).  The dotted line shows the YY-polarization phases 
in the left half of the aperture (the bottom panel in Fig.~11); 
the dashed line shows the YY-polarization
phases in the right half of this aperture.
}
\endfigure

\section{A model for the cross talk}

We have attempted to construct a model for the cross talk.  
A face-on view of the ATCA antenna main reflector, showing the
inter-panel gaps and the intra-panel slits, is in Fig.~13.
All of these gaps and slits are modelled as sources of   
spatially-incoherent broad-band thermal radiation.  
With a pair of antennas 30.6-m apart, pointing nominally 
at an azimuth of 270$^{\circ}$ and elevation of 20$^{\circ}$, and
in a configuration with geometric shadowing, 
the emission from the front side of the 
main reflector surface is  picked up by the receiver of the 
antenna in front after a single reflection off its sub-reflector,
emission from the back surface propagates to the receiver of
the antenna behind after reflections off its main and sub-reflectors.
The model assumes that at any point along the gaps and slits, 
the radiation emerging from the two sides of 
the front antenna main reflector is coherent although spatial
incoherence is assumed along the gaps and slits.

\beginfigure{13}
\centerline{\psfig{file=fig_13.ps,clip=,height=6.0cm,angle=-90}}
\caption{{\bf Figure~13.} 
A face-on view of the ATCA antenna main reflector showing the 
layout of the panels.  The boundaries of the panels are shown with 
thicker lines, these are the lines along which the inter-panel gaps lie;
the intra-panel slits are shown using lighter lines. The plot
axes are in metres.
}
\endfigure

The signal received by the front antenna, from the distributed
source on its main reflector surface, is a constant that is
independent of the antenna pointing.  Radiation 
from the back surface was propagated to the
aperture plane of the antenna behind using a physical optics (PO)
approximation; these signals propagated across space to the
receiver of the antenna behind after reflections off the main 
and sub-reflectors of the antenna behind.  The propagation to the receiver 
of the antenna behind involved an additional reflection as compared 
to the propagation to the receiver in the front antenna and, consequently,
were phase shifted through an additional $\pi$ radians.
The lack of spatial coherence in the emission from the surface was
accounted for in the modelling by computing the correlated response
of the interferometer for each element of the surface emittor
separately and then vectorially summing over the responses.  When both
antennas point towards the same sky direction and their optical
axes are parallel, the path lengths of the GO rays, from every element of 
the surface of the main reflector of the shadowing antenna to the
receivers on the front and back antennas, are identical.  However, in
the GO approximation, the response to this emission will have a phase
of $\pi$ radians owing to the additional reflection off the main 
reflector of the back antenna.  In the PO approximation, the phase
differs from $\pi$ radians even when the optical axes of the
two antennas are parallel and the interferometer visibility phase
depends on the shadowing geometry. 

The emission associated with the gaps and slits was assumed to be
100 per cent polarized on both sides of the surface, the E-field of the
radiation is assumed to be oriented perpendicular to the gap/slit.
The model allowed for a differential weighting to be applied to
the emission from the intra-panel gaps relative to the inter-panel
slits. The model also allows for constant phase terms to be added to
the radiation from the back surface of the main reflector; the
value of this phase is allowed to be different for the gaps and slits.
The weighting may be physically associated with the relative 
intensities of radiation from the gaps with respect to the slits;
the phase terms may be physically
associated with the extra propagation paths for the back propagating 
rays because their direct lines of sight to the back antenna aperture
are blocked by backup structural elements.
The blockage owing to the legs of the tetrapod supporting the
sub-reflector was modelled by omitting those rays that are incident
on any of the two antenna apertures within sectors of $10^{\circ}$
angles centred at the locations of the legs.  Additionally,
the Fresnel diffraction at the tetrapod legs was modelled by
a downweighting of the rays that were incident within $20^{\circ}$
sectors centred at the locations of the legs.
The PO computation was made of the interfometer response, with 
a pair of antennas spaced 30.6-m apart and pointing nominally 
at AZ~=~270$^{\circ}$, EL~=~20$^{\circ}$ and with 
pointing offsets made in azimuth and elevation to the
antenna in front and behind; the offsets were made to only one of the
antennas at any time while the other continued to be pointed at
the nominal position.  The goal of this modelling was to examine whether
the observed scan patterns shown in Figs.~6 \& 7 could be reproduced
with reasonable choices for the parameters of the model.

\beginfigure*{14}
\centerline{\psfig{file=fig_14.ps,clip=,height=9.0cm,angle=-90}}
\caption{{\bf Figure~14.} 
The computed cross-talk amplitudes and phases, in Stokes~I as well
as in the XX and YY polarization products, are shown here
assuming that it arises from 100 per cent linearly polarized 
emission, which is spatially incoherent and distributed along the
gaps and slits in the main reflector surface of the antenna in front.
Two antennas, which are 30.6-m apart E-W, 
point nominally towards AZ~=~$270^{\circ}$
and EL~=~$20^{\circ}$.  The variation in amplitude (continuous 
line) and phase (symbols) of the cross
talk at 4800~MHz are shown as the azimuth of one antenna is offset from
the nominal direction keeping the antenna elevation fixed.  
The panels on the left are for the case where the antenna in front scans in
azimuth; those on the right correspond to the case where the antenna at
the back scans.  The three panels on either side separately show the
Stokes-I visibilities and the XX and YY polarization products.
}
\endfigure
\beginfigure*{15}
\centerline{\psfig{file=fig_15.ps,clip=,height=9.0cm,angle=-90}}
\caption{{\bf Figure~15.} 
As in Fig.~14, with  offsets now applied in elevation instead of azimuth.
Both antennas point towards a constant azimuth of $270^{\circ}$ and
the offsets are made in elevation about the nominal EL~=~$20^{\circ}$.
}
\endfigure

First, the computation confirms that the observed scan patterns 
are broadly reproducible with the source modelled as spatially incoherent
emission distributed over the entire surface of the front antenna
main reflector.  In the shadowed configuration being 
considered here, the individual 
polarization products (XX \& YY) respond to emission associated with 
inter- and intra-panel
circumferential slots in one half of the overlap segment and 
inter-panel radial slots in the other half.  
Introduction of a differential phase term 
between the intra-panel slits -- which are all circumferential -- and the
inter-panel gaps, results in that the polarization products respond
to emission that has a net phase variation across the length of the 
overlap segment.  When one of the antennas in the shadowed
configuration is scanned in azimuth, this phase difference causes 
the XX and YY polarization responses to peak at locations offset
from the nominal pointing.  Indeed, this is what is observed in Fig.~6
and the modelling reveals that a differential phase of about 
$60^{\circ}$ between the gaps and slits reproduces these shifts
as well as the asymmetries in these polarization product scan patterns
at 4800~MHz.  At 8640~MHz, the modelling suggests a phase difference
of about $160^{\circ}$.  The modelling at 4800~MHz required 
that the emission, per unit length, 
from the intra-panel gaps be weighted up relative to that from 
the intra-panel slits: we have used a relative weighting factor of
four for the computation at this frequency.  An overall radial weighting
was implemented by weighting up the components distributed over the
outermost ring of panels by a factor of two as compared to that from
the inner rings.

The computed amplitudes and phases of the cross talk, in Stokes~I and
in the  XX and YY polarization products, versus pointing offsets in 
azimuth and elevation, are shown in Figs.~14 \& 15.
The scan patterns, computed for the model,
agree with the observations in most respects suggesting that we have
a good model for the cross talk.  It is indeed remarkable that we have
been able to successfully model the high-sidelobe azimuthal scans, the
shifts and asymmetries in the XX and YY polarization product azimuth scans,
the elevation scan patterns along with the asymmetries in their
sidelobe structure and the different phase patterns observed 
in the elevation scans when the antenna at the back and front were
separately scanned.  It may be noted here that the PO computation
was important for the modelling (as opposed to a GO approximation); 
in particularly, the PO analysis was essential for reproducing the 
asymmetries in the elevation scan patterns.

There are some aspects of the observed scans that have not been 
reproduced very well and we believe that these 
details might depend on, for example, 
the detailed understanding of the frequency
dependence of the radiative properties of the slots and gaps on the
antenna surface, the
interaction of the back propagating radiation with the backup structural
members and the variation in this interaction with radius. Another
aspect that we have not addressed in the modelling is the possibility
of partial spatial coherence in the emission across the surface of
the main reflector of the antenna in front.  

\section{Commentary}

The cross talk arises because of coherence in the emission, associated
with the inter-panel gaps and intra-panel slits,
from opposite sides of the main reflector of the antenna in front.
The ATCA reflectors are constructed from Aluminium sheets and the skin depth
for cm wavelength radiation is about 1~$\mu$m and any EM field would drop 
exponentially to insignificant values as it propagates through the panels.
Therefore, any cross talk of the kind we observe at the ATCA cannot arise
from radiative propagation through the solid panels and requires a coupling
conduit across the panel surfaces.  Our examination of the nature of the cross
talk, particularly its polarization characteristics, clearly show that
inter-panel gaps and, more surprisingly, intra-panel slits, can serve
as the cause for coherence in the emission from the opposite sides
of the reflector surface. 

The incident EM field from the sky, with an intensity corresponding
to the sky brightness, has a partial spatial coherence 
due to the large-scale brightness temperature variations on the sky.  
The ATCA receiver has a circulator in the signal path between 
the feed horn and the first low
noise amplifier; the circulator has a third port terminated by a 
load that is cooled to cryogenic temperatures and the brightness of
the sub-reflector, as viewed by elements of the main reflector, 
corresponds to the physical temperature of this load.
Elements of the front surfaces of the
panels of the antenna that is in front receive radiation from the 
background sky as well as from the sub-reflector.  
In a configuration where an antenna geometrically shadows
another, the back surfaces of the panels (of the antenna in front) see the
sky reflected off the antenna behind and also see the ground along
other lines of sight.  All these incident fields are potentially
primary sources for the currents on the panel surfaces.
The gaps and slits might be lines along which currents couple
across the panel surfaces causing coherence in surface currents
and, consequently, coherence in the secondary emission
from currents on the panel surfaces. It may be recalled here that the
observations described in section~4.1, showing that the cross-talk
amplitude depends only of the length of the projected shadowed baseline
and may be the same at different antenna elevations and when the
shadowed segments are at different regions of the main reflector surface,
argue against ground re-radiation from the slots as a possible 
mechanism for the cross-talk.

All the panels of the main reflectors are in thermal equilibrium
with the surrounding air that is at ambient temperature.  The panel
surfaces are spatially incoherent emittors of their thermal heat; 
however, because of their high reflectivity the
emissivity of the Al panels is small and corresponds to a brightness
temperature of not more than a few Kelvin.  As discussed above, this
radiation from opposite sides of the solid panel surfaces would not
be expected to be coherent and, therefore, would not result in 
cross talk if they are picked up by different elements of an interferometer.  
Gaps and slits in the panel surfaces would, however, act as slot
antennas and these would emit EM radiation, 
that is coherent, when viewed from 
opposite sides of the surface.  Stochastic voltages are
induced across the gaps and slits corresponding to the physical temperature
of the panels and the efficiency of the slot antenna would depend on
its geometry and its impedance match to free space.

We finally attempt an understanding of the observed strength of the
cross talk.  At 4800~MHz, 
when the projected baseline is 13~m and the antennas are separated by
30.6~m, the cross talk signal is measured by  the ATCA correlator
to have a correlation coefficient of 0.075 per cent.  The geometric mean
system temperature of the antenna pair in this shadowed configuration
is about 60~K implying that the cross talk corresponds to an antenna 
temperature of 45~mK.  In this configuration, where the antennas point towards
an elevation of $25^{\circ}$, the area of geometric shadowing
is 112~m$^{2}$. The area of geometric shadowing is 30 per cent of the
area of the main reflector and this implies that the emission from the 
main reflector surface, which is received by any one of the antennas, 
has a mean brightness temperature of 150~mK.
We may account for this level of radiation, together with the observation 
that the cross-talk amplitude scales inversely with frequency, 
if we assume that the slots act as radiators with 
an effective temperature that is 1--2 per cent of the
physical temperature (about 300~K), and an effective area 
that corresponds to the slot length times a width that is one wavelength.
The first assumption might be related to the 98-99 per cent 
reflectivity of the surface metal, the second aspect might
be related to the fact that a thin slot (or dipole) antenna
has an effective area (as a receiver or radiator) that has 
a width of the order one wavelength. In such a case, the slots
would have an effective area that is about 3 per cent of 
the reflector surface at 4800~MHz.

It is interesting to ask whether all of the additive cross-talk, in
interferometer baselines with geometric shadowing, could
be eliminated by designing the antennas to be made of solid surfaces.
A possible mechanism for cross-talk in this case is scattering/emission
of common-mode power from those parts of the outer rim of the front 
antenna that overlap the antenna behind. However, 
in shadowed configurations, our measurements of the distribution
of the cross-talk across the aperture show no cross-talk component 
associated with the outer rim of the main reflector surface of the 
antenna in front (see Fig.~10 and 11).  

\section{Summary}

The nature of unwanted spurious signals seen in ATCA baselines in
shadowed configurations has been examined.  This cross talk appears
to be a linearly polarized additive component which arrives 
at the receivers of both antennas with almost exactly the delay
that a celestial source would have if it were at the pointing
centre.  

Our examination of the cross talk signal and its variation with
the shadowing configuration indicates that the spurious 
interferometer response is a result of emission from the main
reflector surface of the antenna in front.  Coherence between the
emission from the front and back surfaces of this reflecting surface,
which propagate respectively into the receivers of the antennas in front
and at the back, result in an interferometric response, 
as observed. The coherence 
is owing to the gaps between the panels forming the main reflector
surface as well as slits between the plates that are bonded
together to form the panels.  We conclude that this mode of cross
talk may be avoided in short-spacing interferometers by constructing
the main reflector surfaces as continuous conducting sheets.
In the present case, thin conducting stripes (like Al tape)
covering all the gaps would reduce considerably the cross-talk between
the ATCA dishes in shadowed configurations.
Alternately, this cross talk would be avoided if the backside of the
main reflector is covered using conducting sheets. 

\section*{Acknowledgments}

The Australia Telescope is funded by the Commonwealth of Australia for
operation as a National Facility managed by CSIRO.

\end

%% file: mn.tex
% MN.TEX (Computer Modern version)
%
% plain TeX single / double column macros for the
% Monthly Notices of Royal Astronomical Society
%
% v1.6  (mn.tex)  --- released 18th September 1995 (A. Woollatt)
% v1.5      "     --- released 25th August 1994 (M. Reed)
% v1.4      "     --- released 22nd February 1994
% v1.3  (mnd.tex) --- released 28th November 1992
% v1.26     "     --- released  1st August 1992
% v1.25     "     --- released 25th February 1992
%
% Copyright Cambridge University Press
%
% > Incorporating special symbol code from laa.sty v1.1 (25th Feb 1991)
%   used with the permission of Springer Verlag.
% > Incorporating parts of mssymb.tex (8th July 1987).
% > Incorporating NewFont.sty v ALPHA patchlevel 8 (16th August 1994).
% > Add footlines, add footnotes in double column (18th September
%   1995).

\catcode `\@=11 % @ signs are letters

\def\@version{1.6}
\def\@verdate{18th September 1995}

% Fonts: Computer Modern / Monotype Times (CUP only)
%
% Font family sizes available:
%   8pt, 9pt, 10pt, 11pt, 14pt and 17pt.
%
% Faces available:
%   \rm, math italic, symbol, \it, \bf, \sl, \tt, \sc, \sf, \cal, \em,
%   \mit and \oldstyle.

% define the typeface in use

\newif\ifprod@font

\ifx\@typeface\undefined
  \def\@typeface{Comp. Modern}\prod@fontfalse
\else
  \prod@fonttrue % We want Times
\fi

\def\newfam{\alloc@8\fam\chardef\sixt@@n} % made not outer

\ifprod@font
\font\fiverm=mtr10 at 5pt
\font\fivebf=mtbx10 at 5pt
\font\fiveit=mtti10 at 5pt
\font\fivesl=mtsl10 at 5pt
\font\fivett=cmtt8 at 5pt     \hyphenchar\fivett=-1
\font\fivecsc=mtcsc10 at 5pt
\font\fivesf=mtss10 at 5pt
\font\fivei=mtmi10 at 5pt      \skewchar\fivei='177
\font\fivesy=mtsy10 at 5pt     \skewchar\fivesy='60

\font\sixrm=mtr10 at 6pt
\font\sixbf=mtbx10 at 6pt
\font\sixit=mtti10 at 6pt
\font\sixsl=mtsl10 at 6pt
\font\sixtt=cmtt8 at 6pt      \hyphenchar\sixtt=-1
\font\sixcsc=mtcsc10 at 6pt
\font\sixsf=mtss10 at 6pt
\font\sixi=mtmi10 at 6pt       \skewchar\sixi='177
\font\sixsy=mtsy10 at 6pt      \skewchar\sixsy='60

\font\sevenrm=mtr10 at 7pt
\font\sevenbf=mtbx10 at 7pt
\font\sevenit=mtti10 at 7pt
\font\sevensl=mtsl10 at 7pt
\font\seventt=cmtt8 at 7pt     \hyphenchar\seventt=-1
\font\sevencsc=mtcsc10 at 7pt
\font\sevensf=mtss10 at 7pt
\font\seveni=mtmi10 at 7pt      \skewchar\seveni='177
\font\sevensy=mtsy10 at 7pt     \skewchar\sevensy='60

\font\eightrm=mtr10 at 8pt
\font\eightbf=mtbx10 at 8pt
\font\eightit=mtti10 at 8pt
\font\eighti=mtmi10 at 8pt      \skewchar\eighti='177
\font\eightsy=mtsy10 at 8pt     \skewchar\eightsy='60
\font\eightsl=mtsl10 at 8pt
\font\eighttt=cmtt8             \hyphenchar\eighttt=-1
\font\eightcsc=mtcsc10 at 8pt
\font\eightsf=mtss10 at 8pt

\font\ninerm=mtr10 at 9pt
\font\ninebf=mtbx10 at 9pt
\font\nineit=mtti10 at 9pt
\font\ninei=mtmi10 at 9pt      \skewchar\ninei='177
\font\ninesy=mtsy10 at 9pt     \skewchar\ninesy='60
\font\ninesl=mtsl10 at 9pt
\font\ninett=cmtt9             \hyphenchar\ninett=-1
\font\ninecsc=mtcsc10 at 9pt
\font\ninesf=mtss10 at 9pt

\font\tenrm=mtr10
\font\tenbf=mtbx10
\font\tenit=mtti10
\font\teni=mtmi10		\skewchar\teni='177
\font\tensy=mtsy10		\skewchar\tensy='60
\font\tenex=cmex10
\font\tensl=mtsl10
\font\tentt=cmtt10		\hyphenchar\tentt=-1
\font\tencsc=mtcsc10
\font\tensf=mtss10

\font\elevenrm=mtr10 at 11pt
\font\elevenbf=mtbx10 at 11pt
\font\elevenit=mtti10 at 11pt
\font\eleveni=mtmi10 at 11pt      \skewchar\eleveni='177
\font\elevensy=mtsy10 at 11pt     \skewchar\elevensy='60
\font\elevensl=mtsl10 at 11pt
\font\eleventt=cmtt10 at 11pt     \hyphenchar\eleventt=-1
\font\elevencsc=mtcsc10 at 11pt
\font\elevensf=mtss10 at 11pt

\font\twelverm=mtr10 at 12pt
\font\twelvebf=mtbx10 at 12pt
\font\twelveit=mtti10 at 12pt
\font\twelvesl=mtsl10 at 12pt
\font\twelvett=cmtt12             \hyphenchar\twelvett=-1
\font\twelvecsc=mtcsc10 at 12pt
\font\twelvesf=mtss10 at 12pt
\font\twelvei=mtmi10 at 12pt      \skewchar\twelvei='177
\font\twelvesy=mtsy10 at 12pt     \skewchar\twelvesy='60

\font\fourteenrm=mtr10 at 14pt
\font\fourteenbf=mtbx10 at 14pt
\font\fourteenit=mtti10 at 14pt
\font\fourteeni=mtmi10 at 14pt      \skewchar\fourteeni='177
\font\fourteensy=mtsy10 at 14pt     \skewchar\fourteensy='60
\font\fourteensl=mtsl10 at 14pt
\font\fourteentt=cmtt12 at 14pt     \hyphenchar\fourteentt=-1
\font\fourteencsc=mtcsc10 at 14pt
\font\fourteensf=mtss10 at 14pt

\font\seventeenrm=mtr10 at 17pt
\font\seventeenbf=mtbx10 at 17pt
\font\seventeenit=mtti10 at 17pt
\font\seventeeni=mtmi10 at 17pt      \skewchar\seventeeni='177
\font\seventeensy=mtsy10 at 17pt     \skewchar\seventeensy='60
\font\seventeensl=mtsl10 at 17pt
\font\seventeentt=cmtt12 at 17pt     \hyphenchar\seventeentt=-1
\font\seventeencsc=mtcsc10 at 17pt
\font\seventeensf=mtss10 at 17pt
\else
\font\fiverm=cmr5
\font\fivei=cmmi5             \skewchar\fivei='177
\font\fivesy=cmsy5            \skewchar\fivesy='60
\font\fivebf=cmbx5

\font\sixrm=cmr6
\font\sixi=cmmi6             \skewchar\sixi='177
\font\sixsy=cmsy6            \skewchar\sixsy='60
\font\sixbf=cmbx6

\font\sevenrm=cmr7
\font\sevenit=cmti7
\font\seveni=cmmi7             \skewchar\seveni='177
\font\sevensy=cmsy7            \skewchar\sevensy='60
\font\sevenbf=cmbx7

\font\eightrm=cmr8
\font\eightbf=cmbx8
\font\eightit=cmti8
\font\eighti=cmmi8			\skewchar\eighti='177
\font\eightsy=cmsy8			\skewchar\eightsy='60
\font\eightsl=cmsl8
\font\eighttt=cmtt8			\hyphenchar\eighttt=-1
\font\eightcsc=cmcsc10 at 8pt
\font\eightsf=cmss8

\font\ninerm=cmr9
\font\ninebf=cmbx9
\font\nineit=cmti9
\font\ninei=cmmi9			\skewchar\ninei='177
\font\ninesy=cmsy9			\skewchar\ninesy='60
\font\ninesl=cmsl9
\font\ninett=cmtt9			\hyphenchar\ninett=-1
\font\ninecsc=cmcsc10 at 9pt
\font\ninesf=cmss9

\font\tenrm=cmr10
\font\tenbf=cmbx10
\font\tenit=cmti10
\font\teni=cmmi10		\skewchar\teni='177
\font\tensy=cmsy10		\skewchar\tensy='60
\font\tenex=cmex10
\font\tensl=cmsl10
\font\tentt=cmtt10		\hyphenchar\tentt=-1
\font\tencsc=cmcsc10
\font\tensf=cmss10

\font\elevenrm=cmr10 scaled \magstephalf
\font\elevenbf=cmbx10 scaled \magstephalf
\font\elevenit=cmti10 scaled \magstephalf
\font\eleveni=cmmi10 scaled \magstephalf	\skewchar\eleveni='177
\font\elevensy=cmsy10 scaled \magstephalf	\skewchar\elevensy='60
\font\elevensl=cmsl10 scaled \magstephalf
\font\eleventt=cmtt10 scaled \magstephalf	\hyphenchar\eleventt=-1
\font\elevencsc=cmcsc10 scaled \magstephalf
\font\elevensf=cmss10 scaled \magstephalf

\font\twelverm=cmr10 scaled \magstep1
\font\twelvebf=cmbx10 scaled \magstep1
\font\twelvei=cmmi10 scaled \magstep1      \skewchar\twelvei='177
\font\twelvesy=cmsy10 scaled \magstep1     \skewchar\twelvesy='60

\font\fourteenrm=cmr10 scaled \magstep2
\font\fourteenbf=cmbx10 scaled \magstep2
\font\fourteenit=cmti10 scaled \magstep2
\font\fourteeni=cmmi10 scaled \magstep2		\skewchar\fourteeni='177
\font\fourteensy=cmsy10 scaled \magstep2	\skewchar\fourteensy='60
\font\fourteensl=cmsl10 scaled \magstep2
\font\fourteentt=cmtt10 scaled \magstep2	\hyphenchar\fourteentt=-1
\font\fourteencsc=cmcsc10 scaled \magstep2
\font\fourteensf=cmss10 scaled \magstep2

\font\seventeenrm=cmr10 scaled \magstep3
\font\seventeenbf=cmbx10 scaled \magstep3
\font\seventeenit=cmti10 scaled \magstep3
\font\seventeeni=cmmi10 scaled \magstep3	\skewchar\seventeeni='177
\font\seventeensy=cmsy10 scaled \magstep3	\skewchar\seventeensy='60
\font\seventeensl=cmsl10 scaled \magstep3
\font\seventeentt=cmtt10 scaled \magstep3	\hyphenchar\seventeentt=-1
\font\seventeencsc=cmcsc10 scaled \magstep3
\font\seventeensf=cmss10 scaled \magstep3
\fi

\def\hexnumber#1{\ifcase#1 0\or1\or2\or3\or4\or5\or6\or7\or8\or9\or
  A\or B\or C\or D\or E\or F\fi}

\def\makestrut{%
  \setbox\strutbox=\hbox{%
    \vrule height.7\baselineskip depth.3\baselineskip width \z@}%
}

\def\baselinestretch{1}
\newskip\tmp@bls

\def\b@ls#1{% set baseline using \baselinestretch as a scale factor
  \tmp@bls=#1\relax
  \baselineskip=#1\relax\makestrut
  \normalbaselineskip=\baselinestretch\tmp@bls
  \normalbaselines
}

\def\nostb@ls#1{% set baseline skip ignoring \baselinestretch
  \normalbaselineskip=#1\relax
  \normalbaselines
  \makestrut
}

% families \itfam, \slfam, \bffam, \ttfam defined in PLAIN.
%
% \itfam is \fam4
% \slfam is \fam5
% \bffam is \fam6
% \ttfam is \fam7

\newfam\scfam  % \fam8
\newfam\sffam  % \fam9

\def\mit{\fam\@ne}
\def\cal{\fam\tw@}
\def\em{\ifdim\fontdimen1\font>\z@ \rm\else\it\fi}

\textfont3=\tenex
\scriptfont3=\tenex
\scriptscriptfont3=\tenex

\setbox0=\hbox{\tenex B} \p@renwd=\wd0 % width of the big left (

\def\eightpoint{% 8^6^5 on 10pt
  \def\rm{\fam0\eightrm}%
  \textfont0=\eightrm \scriptfont0=\sixrm \scriptscriptfont0=\fiverm%
  \textfont1=\eighti  \scriptfont1=\sixi  \scriptscriptfont1=\fivei%
  \textfont2=\eightsy \scriptfont2=\sixsy \scriptscriptfont2=\fivesy%
  \textfont\itfam=\eightit\def\it{\fam\itfam\eightit}%
  \ifprod@font
    \scriptfont\itfam=\sixit
      \scriptscriptfont\itfam=\fiveit
  \else
    \scriptfont\itfam=\eightit
      \scriptscriptfont\itfam=\eightit
  \fi
  \textfont\bffam=\eightbf%
    \scriptfont\bffam=\sixbf%
      \scriptscriptfont\bffam=\fivebf%
  \def\bf{\fam\bffam\eightbf}%
  \textfont\slfam=\eightsl\def\sl{\fam\slfam\eightsl}%
  \ifprod@font
    \scriptfont\slfam=\sixsl
      \scriptscriptfont\slfam=\fivesl
  \else
    \scriptfont\slfam=\eightsl
      \scriptscriptfont\slfam=\eightsl
  \fi
  \textfont\ttfam=\eighttt\def\tt{\fam\ttfam\eighttt}%
  \ifprod@font
    \scriptfont\ttfam=\sixtt
      \scriptscriptfont\ttfam=\fivett
  \else
    \scriptfont\ttfam=\eighttt
      \scriptscriptfont\ttfam=\eighttt
  \fi
  \textfont\scfam=\eightcsc\def\sc{\fam\scfam\eightcsc}%
  \ifprod@font
    \scriptfont\scfam=\sixcsc
      \scriptscriptfont\scfam=\fivecsc
  \else
    \scriptfont\scfam=\eightcsc
      \scriptscriptfont\scfam=\eightcsc
  \fi
  \textfont\sffam=\eightsf\def\sf{\fam\sffam\eightsf}%
  \ifprod@font
    \scriptfont\sffam=\sixsf
      \scriptscriptfont\sffam=\fivesf
  \else
    \scriptfont\sffam=\eightsf
      \scriptscriptfont\sffam=\eightsf
  \fi
  \def\oldstyle{\fam\@ne\eighti}%
  \b@ls{10pt}\rm\@viiipt%
}
\def\@viiipt{}

\def\ninepoint{% 9^6^5 on 11pt (two col) / 12 (single col)
  \def\rm{\fam0\ninerm}%
  \textfont0=\ninerm \scriptfont0=\sixrm \scriptscriptfont0=\fiverm%
  \textfont1=\ninei  \scriptfont1=\sixi  \scriptscriptfont1=\fivei%
  \textfont2=\ninesy \scriptfont2=\sixsy \scriptscriptfont2=\fivesy%
  \textfont\itfam=\nineit\def\it{\fam\itfam\nineit}%
  \ifprod@font
    \scriptfont\itfam=\sixit
      \scriptscriptfont\itfam=\fiveit
  \else
    \scriptfont\itfam=\nineit
      \scriptscriptfont\itfam=\nineit
  \fi
  \textfont\bffam=\ninebf%
    \scriptfont\bffam=\sixbf%
      \scriptscriptfont\bffam=\fivebf%
  \def\bf{\fam\bffam\ninebf}%
  \textfont\slfam=\ninesl\def\sl{\fam\slfam\ninesl}%
  \ifprod@font
    \scriptfont\slfam=\sixsl
      \scriptscriptfont\slfam=\fivesl
  \else
    \scriptfont\slfam=\ninesl
      \scriptscriptfont\slfam=\ninesl
  \fi
  \textfont\ttfam=\ninett\def\tt{\fam\ttfam\ninett}%
  \ifprod@font
    \scriptfont\ttfam=\sixtt
      \scriptscriptfont\ttfam=\fivett
  \else
    \scriptfont\ttfam=\ninett
      \scriptscriptfont\ttfam=\ninett
  \fi
  \textfont\scfam=\ninecsc\def\sc{\fam\scfam\ninecsc}%
  \ifprod@font
    \scriptfont\scfam=\sixcsc
      \scriptscriptfont\scfam=\fivecsc
  \else
    \scriptfont\scfam=\ninecsc
      \scriptscriptfont\scfam=\ninecsc
  \fi
  \textfont\sffam=\ninesf\def\sf{\fam\sffam\ninesf}%
  \ifprod@font
    \scriptfont\sffam=\sixsf
      \scriptscriptfont\sffam=\fivesf
  \else
    \scriptfont\sffam=\ninesf
      \scriptscriptfont\sffam=\ninesf
  \fi
  \def\oldstyle{\fam\@ne\ninei}%
  \b@ls{\TextLeading plus \Feathering}\rm\@ixpt%
}
\def\@ixpt{}

\def\tenpoint{% 10^7^5 on 11pt
  \def\rm{\fam0\tenrm}%
  \textfont0=\tenrm \scriptfont0=\sevenrm \scriptscriptfont0=\fiverm%
  \textfont1=\teni  \scriptfont1=\seveni  \scriptscriptfont1=\fivei%
  \textfont2=\tensy \scriptfont2=\sevensy \scriptscriptfont2=\fivesy%
  \textfont\itfam=\tenit\def\it{\fam\itfam\tenit}%
  \ifprod@font
    \scriptfont\itfam=\sevenit
      \scriptscriptfont\itfam=\fiveit
  \else
    \scriptfont\itfam=\tenit
      \scriptscriptfont\itfam=\tenit
  \fi
  \textfont\bffam=\tenbf%
    \scriptfont\bffam=\sevenbf%
      \scriptscriptfont\bffam=\fivebf%
  \def\bf{\fam\bffam\tenbf}%
  \textfont\slfam=\tensl\def\sl{\fam\slfam\tensl}%
  \ifprod@font
    \scriptfont\slfam=\sevensl
      \scriptscriptfont\slfam=\fivesl
  \else
    \scriptfont\slfam=\tensl
      \scriptscriptfont\slfam=\tensl
  \fi
  \textfont\ttfam=\tentt\def\tt{\fam\ttfam\tentt}%
  \ifprod@font
    \scriptfont\ttfam=\seventt
      \scriptscriptfont\ttfam=\fivett
  \else
    \scriptfont\ttfam=\tentt
      \scriptscriptfont\ttfam=\tentt
  \fi
  \textfont\scfam=\tencsc\def\sc{\fam\scfam\tencsc}%
  \ifprod@font
    \scriptfont\scfam=\sevencsc
      \scriptscriptfont\scfam=\fivecsc
  \else
    \scriptfont\scfam=\tencsc
      \scriptscriptfont\scfam=\tencsc
  \fi
  \textfont\sffam=\tensf\def\sf{\fam\sffam\tensf}%
  \ifprod@font
    \scriptfont\sffam=\sevensf
      \scriptscriptfont\sffam=\fivesf
  \else
    \scriptfont\sffam=\tensf
      \scriptscriptfont\sffam=\tensf
  \fi
  \def\oldstyle{\fam\@ne\teni}%
  \b@ls{11pt}\rm\@xpt%
}
\def\@xpt{}

\def\elevenpoint{% 11^8^6 on 13pt
  \def\rm{\fam0\elevenrm}%
  \textfont0=\elevenrm \scriptfont0=\eightrm \scriptscriptfont0=\sixrm%
  \textfont1=\eleveni  \scriptfont1=\eighti  \scriptscriptfont1=\sixi%
  \textfont2=\elevensy \scriptfont2=\eightsy \scriptscriptfont2=\sixsy%
  \textfont\itfam=\elevenit\def\it{\fam\itfam\elevenit}%
  \ifprod@font
    \scriptfont\itfam=\eightit
      \scriptscriptfont\itfam=\sixit
  \else
    \scriptfont\itfam=\elevenit
      \scriptscriptfont\itfam=\elevenit
  \fi
  \textfont\bffam=\elevenbf%
    \scriptfont\bffam=\eightbf%
      \scriptscriptfont\bffam=\sixbf%
  \def\bf{\fam\bffam\elevenbf}%
  \textfont\slfam=\elevensl\def\sl{\fam\slfam\elevensl}%
  \ifprod@font
    \scriptfont\slfam=\eightsl
      \scriptscriptfont\slfam=\sixsl
  \else
    \scriptfont\slfam=\elevensl
      \scriptscriptfont\slfam=\elevensl
  \fi
  \textfont\ttfam=\eleventt\def\tt{\fam\ttfam\eleventt}%
  \ifprod@font
    \scriptfont\ttfam=\eighttt
      \scriptscriptfont\ttfam=\sixtt
  \else
    \scriptfont\ttfam=\eleventt
      \scriptscriptfont\ttfam=\eleventt
  \fi
  \textfont\scfam=\elevencsc\def\sc{\fam\scfam\elevencsc}%
  \ifprod@font
    \scriptfont\scfam=\eightcsc
      \scriptscriptfont\scfam=\sixcsc
  \else
    \scriptfont\scfam=\elevencsc
      \scriptscriptfont\scfam=\elevencsc
  \fi
  \textfont\sffam=\elevensf\def\sf{\fam\sffam\elevensf}%
  \ifprod@font
    \scriptfont\sffam=\eightsf
      \scriptscriptfont\sffam=\sixsf
  \else
    \scriptfont\sffam=\elevensf
      \scriptscriptfont\sffam=\elevensf
  \fi
  \def\oldstyle{\fam\@ne\eleveni}%
  \b@ls{13pt}\rm\@xipt%
}
\def\@xipt{}

\def\fourteenpoint{% 14^10^7 on 17pt
  \def\rm{\fam0\fourteenrm}%
  \textfont0\fourteenrm  \scriptfont0\tenrm  \scriptscriptfont0\sevenrm%
  \textfont1\fourteeni   \scriptfont1\teni   \scriptscriptfont1\seveni%
  \textfont2\fourteensy  \scriptfont2\tensy  \scriptscriptfont2\sevensy%
  \textfont\itfam=\fourteenit\def\it{\fam\itfam\fourteenit}%
  \ifprod@font
    \scriptfont\itfam=\tenit
      \scriptscriptfont\itfam=\sevenit
  \else
    \scriptfont\itfam=\fourteenit
      \scriptscriptfont\itfam=\fourteenit
  \fi
  \textfont\bffam=\fourteenbf%
    \scriptfont\bffam=\tenbf%
      \scriptscriptfont\bffam=\sevenbf%
  \def\bf{\fam\bffam\fourteenbf}%
  \textfont\slfam=\fourteensl\def\sl{\fam\slfam\fourteensl}%
  \ifprod@font
    \scriptfont\slfam=\tensl
      \scriptscriptfont\slfam=\sevensl
  \else
    \scriptfont\slfam=\fourteensl
      \scriptscriptfont\slfam=\fourteensl
  \fi
  \textfont\ttfam=\fourteentt\def\tt{\fam\ttfam\fourteentt}%
  \ifprod@font
    \scriptfont\ttfam=\tentt
      \scriptscriptfont\ttfam=\seventt
  \else
    \scriptfont\ttfam=\fourteentt
      \scriptscriptfont\ttfam=\fourteentt
  \fi
  \textfont\scfam=\fourteencsc\def\sc{\fam\scfam\fourteencsc}%
  \ifprod@font
    \scriptfont\scfam=\tencsc
      \scriptscriptfont\scfam=\sevencsc
  \else
    \scriptfont\scfam=\fourteencsc
      \scriptscriptfont\scfam=\fourteencsc
  \fi
  \textfont\sffam=\fourteensf\def\sf{\fam\sffam\fourteensf}%
  \ifprod@font
    \scriptfont\sffam=\tensf
      \scriptscriptfont\sffam=\sevensf
  \else
    \scriptfont\sffam=\fourteensf
      \scriptscriptfont\sffam=\fourteensf
  \fi
  \def\oldstyle{\fam\@ne\fourteeni}%
  \b@ls{17pt}\rm\@xivpt%
}
\def\@xivpt{}

\def\seventeenpoint{% 17^12^10 on 20pt
  \def\rm{\fam0\seventeenrm}%
  \textfont0\seventeenrm  \scriptfont0\twelverm  \scriptscriptfont0\tenrm%
  \textfont1\seventeeni   \scriptfont1\twelvei   \scriptscriptfont1\teni%
  \textfont2\seventeensy  \scriptfont2\twelvesy  \scriptscriptfont2\tensy%
  \textfont\itfam=\seventeenit\def\it{\fam\itfam\seventeenit}%
  \ifprod@font
    \scriptfont\itfam=\twelveit
      \scriptscriptfont\itfam=\tenit
  \else
    \scriptfont\itfam=\seventeenit
      \scriptscriptfont\itfam=\seventeenit
  \fi
  \textfont\bffam=\seventeenbf%
    \scriptfont\bffam=\twelvebf%
      \scriptscriptfont\bffam=\tenbf%
  \def\bf{\fam\bffam\seventeenbf}%
  \textfont\slfam=\seventeensl\def\sl{\fam\slfam\seventeensl}%
  \ifprod@font
    \scriptfont\slfam=\twelvesl
      \scriptscriptfont\slfam=\tensl
  \else
    \scriptfont\slfam=\seventeensl
      \scriptscriptfont\slfam=\seventeensl
  \fi
  \textfont\ttfam=\seventeentt\def\tt{\fam\ttfam\seventeentt}%
  \ifprod@font
    \scriptfont\ttfam=\twelvett
      \scriptscriptfont\ttfam=\tentt
  \else
    \scriptfont\ttfam=\seventeentt
      \scriptscriptfont\ttfam=\seventeentt
  \fi
  \textfont\scfam=\seventeencsc\def\sc{\fam\scfam\seventeencsc}%
  \ifprod@font
    \scriptfont\scfam=\twelvecsc
      \scriptscriptfont\scfam=\tencsc
  \else
    \scriptfont\scfam=\seventeencsc
      \scriptscriptfont\scfam=\seventeencsc
  \fi
  \textfont\sffam=\seventeensf\def\sf{\fam\sffam\seventeensf}%
  \ifprod@font
    \scriptfont\sffam=\twelvesf
      \scriptscriptfont\sffam=\tensf
  \else
    \scriptfont\sffam=\seventeensf
      \scriptscriptfont\sffam=\seventeensf
  \fi
  \def\oldstyle{\fam\@ne\seventeeni}%
  \b@ls{20pt}\rm\@xviipt%
}
\def\@xviipt{}

\lineskip=1pt      \normallineskip=\lineskip
\lineskiplimit=\z@ \normallineskiplimit=\lineskiplimit

% BOLD MATH SYMBOLS

% Make \, work in non-math mode
\def\,{\relax\ifmmode \mskip\thinmuskip\else \thinspace\fi}
\let\protect=\relax

\long\def\@ifundefined#1#2#3{\expandafter\ifx\csname
  #1\endcsname\relax#2\else#3\fi}

%%%%%%%%%%%%%%%%%%%%%%%%%%%%%%%%%%%%%%%%%

% NewFont.sty: ALPHA VERSION patchlevel 8, 16th August 1994, M. Reed

% \addtom@thgroup{math font loading info}
% Adds to internal \math@groups definition, which is executed at the end
% of each size changing command. It is called by \NewSymbolFont.

\newtoks\math@groups \math@groups={}
\def\addtom@thgroup#1#2{#1\expandafter{\the#1#2}} %  \mac={new\the\mac}

% Make TeX change the values of \s@ze, \ss@ze, \sss@ze when \@npt is
% executed. This makes it possible for math characters to be loaded
% at the correct size automatically when the size is changed.

% \addtosizeh@ok{x}{10}{7}{5}

\def\addtosizeh@ok#1#2#3#4{%
  \expandafter\def\csname @#1pt\endcsname{%
    \def\s@ze{#2}\def\ss@ze{#3}\def\sss@ze{#4}\the\math@groups%
  }%
}

% \resetsizehook allows the size parameters to be reset after \addtosizeh@ok
% has been called (it re-defines \@npt).
% e.g JFM which requires \xpt to have 10.5pt instead of 10pt.
% Note: \resetsizehook must be used in the preamble BEFORE any
% \New... commands.

% e.g. \resetsizehook{x}{10.5}{7}{5}

\let\resetsizehook=\addtosizeh@ok

% Standard LaTeX sizes

\ifprod@font
%  \addtosizeh@ok{v}    {5} {5}  {5}
%  \addtosizeh@ok{vi}   {6} {6}  {6}
%  \addtosizeh@ok{vii}  {7} {6}  {5}
  \addtosizeh@ok{viii} {8} {6}  {5}
  \addtosizeh@ok{ix}   {9} {6}  {5}
  \addtosizeh@ok{x}    {10}{7}  {5}
  \addtosizeh@ok{xi}   {11}{8}  {6}
%  \addtosizeh@ok{xii}  {12}{8}  {6}
  \addtosizeh@ok{xiv}  {14}{10} {7}
  \addtosizeh@ok{xvii} {17}{12}{10}
%  \addtosizeh@ok{xx}   {20}{14}{12}
%  \addtosizeh@ok{xxv}  {25}{20}{17}
\else
%  \addtosizeh@ok{v}    {5}     {5}     {5}
%  \addtosizeh@ok{vi}   {6}     {6}     {6}
%  \addtosizeh@ok{vii}  {7}     {6}     {5}
  \addtosizeh@ok{viii} {8}     {6}     {5}
  \addtosizeh@ok{ix}   {9}     {6}     {5}
  \addtosizeh@ok{x}    {10}    {7}     {5}
  \addtosizeh@ok{xi}   {10.95} {8}     {6}
%  \addtosizeh@ok{xii}  {12}    {8}     {6}
  \addtosizeh@ok{xiv}  {14.4}  {10}    {7}
  \addtosizeh@ok{xvii} {17.28} {12}    {10}
%  \addtosizeh@ok{xx}   {20.74} {14.4}  {12}
%  \addtosizeh@ok{xxv}  {24.88} {20.74} {17.28}
\fi

\def\get@font#1#2#3{%
  \edef\fonts@ze{\romannumeral#3}%         10 -> x
  \edef\fontn@me{\fonts@ze#1}%             AMSa -> xAMSa
  \@ifundefined{\fontn@me}%
    {%%\typeout{defining \fontn@me}%
     \global\expandafter\font\csname \fontn@me\endcsname=#2 at #3pt}%
    {}%
}

\def\ass@tfont#1#2{%
  \xdef\fam@name{\csname #1\endcsname}%
  \xdef\font@name{\csname #2\endcsname}%
  \let\textfont@name\font@name
  \textfont\fam@name\textfont@name
}

\def\ass@sfont#1#2{%
  \xdef\fam@name{\csname #1\endcsname}%
  \xdef\font@name{\csname #2\endcsname}%
  \let\textfont@name\font@name
  \scriptfont\fam@name\textfont@name
}

\def\ass@ssfont#1#2{%
  \xdef\fam@name{\csname #1\endcsname}%
  \xdef\font@name{\csname #2\endcsname}%
  \let\textfont@name\font@name
  \scriptscriptfont\fam@name\textfont@name
}

%                fam name  base font  (allocates a \newfam)
% \NewSymbolFont {AMSa}    {mtxm10}

\def\NewSymbolFont#1#2{%
  \expandafter\ifx\csname sym#1fam\endcsname\relax % if not defined
    \expandafter\newfam\csname sym#1fam\endcsname
    \expandafter\edef\csname sym#1fam\endcsname{\the\allocationnumber}%
    \addtom@thgroup\math@groups{%
      \get@font{#1}{#2}{\s@ze}%
      \ass@tfont{sym#1fam}{\fontn@me}%
      \get@font{#1}{#2}{\ss@ze}%
      \ass@sfont{sym#1fam}{\fontn@me}%
      \get@font{#1}{#2}{\sss@ze}%
      \ass@ssfont{sym#1fam}{\fontn@me}%
    }%
  \else
    \errmessage{Family `#1' already defined}%
  \fi
}

%                symbol         type fam    pos (hex)
% \NewMathSymbol {\blacksquare} {0}  {AMSa} {04}

\def\NewMathSymbol#1#2#3#4{%
  \edef\f@mly{\expandafter\hexnumber{\csname sym#3fam\endcsname}}%
  \mathchardef#1="#2\f@mly#4\relax
}

%                  macro name  type  fam1   pos  fam2   pos
% \NewMathDelimiter{\ulcorner} {4}   {AMSa} {70} {AMSb} {70}

\newif\ifd@f

\def\NewMathDelimiter#1#2#3#4#5#6{%
  \d@ftrue
  \expandafter\ifx\csname sym#3fam\endcsname\relax
    \d@ffalse \errmessage{Family `#3' is not defined}%
  \fi
  \expandafter\ifx\csname sym#5fam\endcsname\relax
    \d@ffalse \errmessage{Family `#5' is not defined}%
  \fi
  \ifd@f
    \edef\f@mly{\expandafter\hexnumber{\csname sym#3fam\endcsname}}%
    \edef\f@mlytw@{\expandafter\hexnumber{\csname sym#5fam\endcsname}}%
    \xdef#1{\delimiter"#2\f@mly #4\f@mlytw@ #6\relax}%
  \fi
}

%                  macro name  base font  skewchar setting e.g '60 (octal)
% \NewMathAlphabet {mathbssi}  {mtmisb10} {}

\def\setboxz@h{\setbox\z@\hbox}
\def\wdz@{\wd\z@}
\def\boxz@{\box\z@}
\def\setbox@ne{\setbox\@ne}
\def\wd@ne{\wd\@ne}

\def\math@atom#1#2{%
   \binrel@{#1}\binrel@@{#2}}
\def\binrel@#1{\setboxz@h{\thinmuskip0mu
  \medmuskip\m@ne mu\thickmuskip\@ne mu$#1\m@th$}%
 \setbox@ne\hbox{\thinmuskip0mu\medmuskip\m@ne mu\thickmuskip
  \@ne mu${}#1{}\m@th$}%
 \setbox\tw@\hbox{\hskip\wd@ne\hskip-\wdz@}}
\def\binrel@@#1{\ifdim\wd2<\z@\mathbin{#1}\else\ifdim\wd\tw@>\z@
 \mathrel{#1}\else{#1}\fi\fi}

\def\m@thit{1}

\def\set@skchar#1{\global\expandafter\skewchar
  \csname\fontn@me\endcsname=#1\relax}

\def\NewMathAlphabet#1#2#3{%
  \def\tst{#3}%
  \ifx\tst\empty\else % if a \skewchar setting is present
    \expandafter\gdef\csname #1@sc\endcsname{}%  \def\cmd@sc{..}
  \fi
  \expandafter\def\csname #1\endcsname{%  \def\cmd{\protect\@cmd}
    \protect\csname @#1\endcsname}%
  \expandafter\def\csname @#1\endcsname##1{%  \def\@cmd{..}
    {%
    \begingroup
      \get@font{#1}{#2}{\s@ze}%
      \@ifundefined{#1@sc}{}{\set@skchar{#3}}%
      \ass@tfont{m@thit}{\fontn@me}%
      \get@font{#1}{#2}{\ss@ze}%
      \@ifundefined{#1@sc}{}{\set@skchar{#3}}%
      \ass@sfont{m@thit}{\fontn@me}%
      \get@font{#1}{#2}{\sss@ze}%
      \@ifundefined{#1@sc}{}{\set@skchar{#3}}%
      \ass@ssfont{m@thit}{\fontn@me}%
      \math@atom{##1}{%
      \mathchoice%
        {\hbox{$\m@th\displaystyle##1$}}%
        {\hbox{$\m@th\textstyle##1$}}%
        {\hbox{$\m@th\scriptstyle##1$}}%
        {\hbox{$\m@th\scriptscriptstyle##1$}}}%
    \endgroup
    }%
  }%
}

%                  macro name  base font  hyphenchar setting e.g -1 (off)
% \NewTextAlphabet {textbfit}  {mtbxti10} {}

% save a family if \NewTextAlphabet is not used.
\newif\iffirstta  \firsttatrue

\def\set@hchar#1{\global\expandafter\hyphenchar
  \csname\fontn@me\endcsname=#1\relax}

\def\NewTextAlphabet#1#2#3{%
  \iffirstta
    \global\firsttafalse
    \newfam\scratchfam
    \edef\scrt@fam{\the\allocationnumber}%
  \fi
  \def\tst{#3}%
  \ifx\tst\empty\else % if a \hyphenchar setting is required
    \expandafter\gdef\csname #1@hc\endcsname{}%  \def\cmd@sc{..}
  \fi
  \expandafter\def\csname #1\endcsname{%  \def\cmd{\protect\t@cmd}
    \protect\csname t@#1\endcsname}%
  \long\expandafter\def\csname t@#1\endcsname##1{%  \def\t@cmd{..}
    \ifmmode
      \typeout{Warning: do not use \expandafter\string\csname #1\endcsname
        \space in math mode}\fi%
    {%
      \get@font{#1}{#2}{\s@ze}\let\t@xtfnt=\fontn@me\relax
      \@ifundefined{#1@hc}{}{\set@hchar{#3}}%
      \ass@tfont{scrt@fam}{\fontn@me}%
      \get@font{#1}{#2}{\ss@ze}%
      \@ifundefined{#1@hc}{}{\set@hchar{#3}}%
      \ass@sfont{scrt@fam}{\fontn@me}%
      \get@font{#1}{#2}{\sss@ze}%
      \@ifundefined{#1@hc}{}{\set@hchar{#3}}%
      \ass@ssfont{scrt@fam}{\fontn@me}%
      \fam\scratchfam\csname\t@xtfnt\endcsname
    ##1%
    }%
  }%
  \expandafter\def\csname #1shape%  \def\cmdshape{\protect\@cmdshape}
    \endcsname{\protect\csname @#1shape\endcsname}%
  \expandafter\def\csname @#1shape\endcsname{%  \def\@cmdshape
    \ifmmode
      \typeout{Warning: do not use \expandafter\string\csname
        #1shape\endcsname \space in math mode}\fi
      \get@font{#1}{#2}{\s@ze}\let\t@xtfnt=\fontn@me\relax
      \@ifundefined{#1@hc}{}{\set@hchar{#3}}%
      \ass@tfont{scrt@fam}{\fontn@me}%
      \get@font{#1}{#2}{\ss@ze}%
      \@ifundefined{#1@hc}{}{\set@hchar{#3}}%
      \ass@sfont{scrt@fam}{\fontn@me}%
      \get@font{#1}{#2}{\sss@ze}%
      \@ifundefined{#1@hc}{}{\set@hchar{#3}}%
      \ass@ssfont{scrt@fam}{\fontn@me}%
      \fam\scratchfam\csname\t@xtfnt\endcsname
  }%
}

% \bmath{math text}

\ifprod@font
  \def\math@itfnt{mtmib10}
  \def\math@syfnt{mtbsy10}
\else
  \def\math@itfnt{cmmib10}
  \def\math@syfnt{cmbsy10}
\fi

\def\m@thsy{2}

\def\bmath{\protect\@bmath}
\def\@bmath#1{%
  {%
  \begingroup
    \get@font{mthit}{\math@itfnt}{\s@ze}\set@skchar{'177}%
    \ass@tfont{m@thit}{\fontn@me}%
    \get@font{mthit}{\math@itfnt}{\ss@ze}\set@skchar{'177}%
    \ass@sfont{m@thit}{\fontn@me}%
    \get@font{mthit}{\math@itfnt}{\sss@ze}\set@skchar{'177}%
    \ass@ssfont{m@thit}{\fontn@me}%
    \get@font{mthsy}{\math@syfnt}{\s@ze}\set@skchar{'60}%
    \ass@tfont{m@thsy}{\fontn@me}%
    \get@font{mthsy}{\math@syfnt}{\ss@ze}\set@skchar{'60}%
    \ass@sfont{m@thsy}{\fontn@me}%
    \get@font{mthsy}{\math@syfnt}{\sss@ze}\set@skchar{'60}%
    \ass@ssfont{m@thsy}{\fontn@me}%
    \math@atom{#1}{%
    \mathchoice%
      {\hbox{$\m@th\displaystyle#1$}}%
      {\hbox{$\m@th\textstyle#1$}}%
      {\hbox{$\m@th\scriptstyle#1$}}%
      {\hbox{$\m@th\scriptscriptstyle#1$}}}%
  \endgroup
  }%
}

%%%%%%%%%%%%%%%%%%%%%%%%%%%%%%%%%%%%%%%%%

% Astronomy and Astrophysics symbol macros

\def\diameter{{\ifmmode\mathchoice
{\ooalign{\hfil\hbox{$\displaystyle/$}\hfil\crcr
{\hbox{$\displaystyle\mathchar"20D$}}}}
{\ooalign{\hfil\hbox{$\textstyle/$}\hfil\crcr
{\hbox{$\textstyle\mathchar"20D$}}}}
{\ooalign{\hfil\hbox{$\scriptstyle/$}\hfil\crcr
{\hbox{$\scriptstyle\mathchar"20D$}}}}
{\ooalign{\hfil\hbox{$\scriptscriptstyle/$}\hfil\crcr
{\hbox{$\scriptscriptstyle\mathchar"20D$}}}}
\else{\ooalign{\hfil/\hfil\crcr\mathhexbox20D}}%
\fi}}

\def\sq{\ifmmode\squareforqed\else{\unskip\nobreak\hfil
\penalty50\hskip1em\null\nobreak\hfil\squareforqed
\parfillskip=0pt\finalhyphendemerits=0\endgraf}\fi}
\def\squareforqed{\hbox{\rlap{$\sqcap$}$\sqcup$}}

% Simulated Blackboard Bold symbols

\def\bbbc{{\mathchoice {\setbox0=\hbox{$\displaystyle\rm C$}\hbox{\hbox
to0pt{\kern0.4\wd0\vrule height0.9\ht0\hss}\box0}}
{\setbox0=\hbox{$\textstyle\rm C$}\hbox{\hbox
to0pt{\kern0.4\wd0\vrule height0.9\ht0\hss}\box0}}
{\setbox0=\hbox{$\scriptstyle\rm C$}\hbox{\hbox
to0pt{\kern0.4\wd0\vrule height0.9\ht0\hss}\box0}}
{\setbox0=\hbox{$\scriptscriptstyle\rm C$}\hbox{\hbox
to0pt{\kern0.4\wd0\vrule height0.9\ht0\hss}\box0}}}}
\def\bbbq{{\mathchoice {\setbox0=\hbox{$\displaystyle\rm
Q$}\hbox{\raise
0.15\ht0\hbox to0pt{\kern0.4\wd0\vrule height0.8\ht0\hss}\box0}}
{\setbox0=\hbox{$\textstyle\rm Q$}\hbox{\raise
0.15\ht0\hbox to0pt{\kern0.4\wd0\vrule height0.8\ht0\hss}\box0}}
{\setbox0=\hbox{$\scriptstyle\rm Q$}\hbox{\raise
0.15\ht0\hbox to0pt{\kern0.4\wd0\vrule height0.7\ht0\hss}\box0}}
{\setbox0=\hbox{$\scriptscriptstyle\rm Q$}\hbox{\raise
0.15\ht0\hbox to0pt{\kern0.4\wd0\vrule height0.7\ht0\hss}\box0}}}}
\def\bbbt{{\mathchoice {\setbox0=\hbox{$\displaystyle\rm
T$}\hbox{\hbox to0pt{\kern0.3\wd0\vrule height0.9\ht0\hss}\box0}}
{\setbox0=\hbox{$\textstyle\rm T$}\hbox{\hbox
to0pt{\kern0.3\wd0\vrule height0.9\ht0\hss}\box0}}
{\setbox0=\hbox{$\scriptstyle\rm T$}\hbox{\hbox
to0pt{\kern0.3\wd0\vrule height0.9\ht0\hss}\box0}}
{\setbox0=\hbox{$\scriptscriptstyle\rm T$}\hbox{\hbox
to0pt{\kern0.3\wd0\vrule height0.9\ht0\hss}\box0}}}}
\def\bbbs{{\mathchoice
{\setbox0=\hbox{$\displaystyle     \rm S$}\hbox{\raise0.5\ht0\hbox
to0pt{\kern0.35\wd0\vrule height0.45\ht0\hss}\hbox
to0pt{\kern0.55\wd0\vrule height0.5\ht0\hss}\box0}}
{\setbox0=\hbox{$\textstyle        \rm S$}\hbox{\raise0.5\ht0\hbox
to0pt{\kern0.35\wd0\vrule height0.45\ht0\hss}\hbox
to0pt{\kern0.55\wd0\vrule height0.5\ht0\hss}\box0}}
{\setbox0=\hbox{$\scriptstyle      \rm S$}\hbox{\raise0.5\ht0\hbox
to0pt{\kern0.35\wd0\vrule height0.45\ht0\hss}\raise0.05\ht0\hbox
to0pt{\kern0.5\wd0\vrule height0.45\ht0\hss}\box0}}
{\setbox0=\hbox{$\scriptscriptstyle\rm S$}\hbox{\raise0.5\ht0\hbox
to0pt{\kern0.4\wd0\vrule height0.45\ht0\hss}\raise0.05\ht0\hbox
to0pt{\kern0.55\wd0\vrule height0.45\ht0\hss}\box0}}}}
\def\bbbz{{\mathchoice {\hbox{$\sf\textstyle Z\kern-0.4em Z$}}
{\hbox{$\sf\textstyle Z\kern-0.4em Z$}}
{\hbox{$\sf\scriptstyle Z\kern-0.3em Z$}}
{\hbox{$\sf\scriptscriptstyle Z\kern-0.2em Z$}}}}

% NUMBER THE DESIGN ELEMENTS

\def\Nulle{0} % null element
\def\Afe{1}   % author affiliation
\def\Hae{2}   % heading A
\def\Hbe{3}   % heading B
\def\Hce{4}   % heading C
\def\Hde{5}   % heading D

% TEMPORARY REGISTERS

\newcount\LastMac       \LastMac=\Nulle

\newskip\half      \half=5.5pt plus 1.5pt minus 2.25pt
\newskip\one       \one=11pt plus 3pt minus 5.5pt
\newskip\onehalf   \onehalf=16.5pt plus 5.5pt minus 8.25pt
\newskip\two       \two=22pt plus 5.5pt minus 11pt

\def\Half{\addvspace{\half}}
\def\One{\addvspace{\one}}
\def\OneHalf{\addvspace{\onehalf}}
\def\Two{\addvspace{\two}}

\def\Raggedright{% set lines unjustified
  \rightskip=\z@ plus \hsize\relax
}

\def\Fullout{% set lines justified
  \rightskip=\z@\relax
}

\def\Hang#1#2{% set hanging indentation
  \hangindent=#1%
  \hangafter=#2\relax
}

% Pagestyles

\newif\ifsp@page
\def\pagestyle#1{\csname ps@#1\endcsname}
\def\thispagestyle#1{\global\sp@pagetrue\gdef\sp@type{#1}}

\def\ps@titlepage{%
  \def\@oddhead{\eightpoint\noindent \the\CatchLine
    \ifprod@font\else\qquad Printed\ \today\qquad
      (MN plain \TeX\ macros\ v\@version)\fi \hfil}%
  \let\@evenhead=\@oddhead
  \def\@oddfoot{\eightpoint\copyright\ \@pubyear\ RAS\hfil}%
  \def\@evenfoot{\hfil\eightpoint\noindent\copyright\ \@pubyear\ RAS}%
}

\def\ps@headings{%
  \def\@oddhead{\elevenpoint\it\noindent
    \hfill\the\RightHeader\hskip1.5em\rm\folio}%
  \def\@evenhead{\elevenpoint\noindent
    \folio\hskip1.5em\it\the\LeftHeader\hfill}%
  \def\@oddfoot{\eightpoint\noindent\copyright\ \@pubyear\ RAS,
    MNRAS {\bf \@volume}, \@pagerange\hfil}%
  \def\@evenfoot{\hfil\eightpoint\copyright\ \@pubyear\ RAS,
    MNRAS {\bf \@volume}, \@pagerange}%
}

\def\ps@plate{%
  \def\@oddhead{\eightpoint\noindent\plt@cap\hfil}%
  \def\@evenhead{\eightpoint\noindent\plt@cap\hfil}%
  \def\@oddfoot{\eightpoint\noindent\copyright\ \@pubyear\ RAS,
    MNRAS {\bf \@volume}, \@pagerange\hfil}%
  \def\@evenfoot{\hfil\eightpoint\copyright\ \@pubyear\ RAS,
    MNRAS {\bf \@volume}, \@pagerange}%
}

% DESIGN ELEMENT DEFINITIONS

% Article opening

\def\title#1{% article title
  \bgroup
    \vbox to 8pt{\vss}%
    \seventeenpoint
    \Raggedright
    \noindent \strut{\bf #1}\par
  \egroup
}

\def\author#1{% article author(s)
  \bgroup
    \ifnum\LastMac=\Afe \OneHalf\else \vskip 21pt\fi
    \fourteenpoint
    \Raggedright
    \noindent \strut #1\par
    \vskip 3pt%
  \egroup
}

\def\affiliation#1{% author(s) affiliation
  \bgroup
    \vskip -4pt%
    \eightpoint
    \Raggedright
    \noindent \strut {\it #1}\par
  \egroup
  \LastMac=\Afe\relax
}

\def\acceptedline#1{% acceptance date
  \bgroup
    \Two
    \eightpoint
    \Raggedright
    \noindent \strut #1\par
  \egroup
}

\long\def\abstract#1{%
  \bgroup
    \vskip 20pt%
    \leftskip 11pc\rightskip\z@
    \noindent{\ninebf ABSTRACT}\par
    \tenpoint
    \Fullout
    \noindent #1\par
  \egroup
}

\long\def\keywords#1{% keywords
  \bgroup
    \Half
    \leftskip 11pc\rightskip\z@
    \tenpoint
    \Fullout
    \noindent\hbox{\bf Key words:}\ #1\par
  \egroup
}

% The \maketitle macro ensures that the two spanning material appears
% at the top of the first page, and that it has two lines of space
% underneath it. If you forget this in you input, no output will be produced.
% The \BeginOpening (alias \begintopmatter) macro should be called at the
% very start of the input file, so that it is in force when the document
% starts. This ensures that when \maketitle is called that the group is
% closed, and the material gets printed.

\def\maketitle{%
  \EndOpening
  \ifsinglecol \else \MakePage\fi
}

% Page offset

\def\pageoffset#1#2{\hoffset=#1\relax\voffset=#2\relax}

% Counter setup

\def\@nameuse#1{\csname #1\endcsname}
\def\arabic#1{\@arabic{\@nameuse{#1}}}
\def\alph#1{\@alph{\@nameuse{#1}}}
\def\Alph#1{\@Alph{\@nameuse{#1}}}
\def\@arabic#1{\number #1}
\def\@Alph#1{\ifcase#1\or A\or B\or C\or D\else\@Ialph{#1}\fi}
\def\@Ialph#1{\ifcase#1\or \or \or \or \or E\or F\or G\or H\or I\or J\or
   K\or L\or M\or N\or O\or P\or Q\or R\or S\or T\or U\or V\or W\or X\or
   Y\or Z\else\errmessage{Counter out of range}\fi}
\def\@alph#1{\ifcase#1\or a\or b\or c\or d\else\@ialph{#1}\fi}
\def\@ialph#1{\ifcase#1\or \or \or \or \or e\or f\or g\or h\or i\or j\or
   k\or l\or m\or n\or o\or p\or q\or r\or s\or t\or u\or v\or w\or x\or y\or
   z\else\errmessage{Counter out of range}\fi}

% Equation auto-numbering

\newcount\Eqnno
\newcount\SubEqnno

\def\theeq{\arabic{Eqnno}}
\def\thesubeq{\alph{SubEqnno}}

\def\stepeq{\relax
  \global\SubEqnno \z@
  \global\advance\Eqnno \@ne\relax
  {\rm (\theeq)}%
}

\def\startsubeq{\relax
  \global\SubEqnno \z@
  \global\advance\Eqnno \@ne\relax
  \stepsubeq
}

\def\stepsubeq{\relax
  \global\advance\SubEqnno \@ne\relax
  {\rm (\theeq\thesubeq)}%
}

% Headings

\newcount\Sec        %  heading auto number counters
\newcount\SecSec
\newcount\SecSecSec

\def\thesection{\arabic{Sec}}
\def\thesubsection{\thesection.\arabic{SecSec}}
\def\thesubsubsection{\thesubsection.\arabic{SecSecSec}}

\Sec=\z@

\def\:{\let\@sptoken= } \:  % this makes \@sptoken a space token 
\def\:{\@xifnch} \expandafter\def\: {\futurelet\@tempc\@ifnch}

\def\@ifnextchar#1#2#3{%
  \let\@tempMACe #1%
  \def\@tempMACa{#2}%
  \def\@tempMACb{#3}%
  \futurelet \@tempMACc\@ifnch%
}

\def\@ifnch{%
\ifx \@tempMACc \@sptoken%
  \let\@tempMACd\@xifnch%
\else%
  \ifx \@tempMACc \@tempMACe%
    \let\@tempMACd\@tempMACa%
  \else%
    \let\@tempMACd\@tempMACb%
  \fi%
\fi%
\@tempMACd%
}

\def\@ifstar#1#2{\@ifnextchar *{\def\@tempMACa*{#1}\@tempMACa}{#2}}

\newskip\@tempskipb

\def\addvspace#1{%
  \ifvmode\else \endgraf\fi%
  \ifdim\lastskip=\z@%
    \vskip #1\relax%
  \else%
    \@tempskipb#1\relax\@xaddvskip%
  \fi%
}

\def\@xaddvskip{%
  \ifdim\lastskip<\@tempskipb%
    \vskip-\lastskip%
    \vskip\@tempskipb\relax%
  \else%
    \ifdim\@tempskipb<\z@%
      \ifdim\lastskip<\z@ \else%
        \advance\@tempskipb\lastskip%
        \vskip-\lastskip\vskip\@tempskipb%
      \fi%
    \fi%
  \fi%
}

\newskip\@tmpSKIP

\def\addpen#1{%
  \ifvmode
    \if@nobreak
    \else
      \ifdim\lastskip=\z@
        \penalty#1\relax
      \else
        \@tmpSKIP=\lastskip
        \vskip -\lastskip
        \penalty#1\vskip\@tmpSKIP
      \fi
    \fi
  \fi
}

\newcount\@clubpen   \@clubpen=\clubpenalty
\newif\if@nobreak    \@nobreakfalse

\def\@noafterindent{%
  \global\@nobreaktrue
  \everypar{\if@nobreak
              \global\@nobreakfalse
              \clubpenalty \@M
              {\setbox\z@\lastbox}%
              \LastMac=\Nulle\relax%
            \else
              \clubpenalty \@clubpen
              \everypar{}%
            \fi}%
}

\newcount\gds@cbrk   \gds@cbrk=-300

\def\@nohdbrk{\interlinepenalty \@M\relax}

\let\@par=\par
\def\@restorepar{\def\par{\@par}}

\newif\if@endpe   \@endpefalse
 
\def\@doendpe{\@endpetrue \@nobreakfalse \LastMac=\Nulle\relax%
     \def\par{\@restorepar\everypar{}\par\@endpefalse}%
              \everypar{\setbox\z@\lastbox\everypar{}\@endpefalse}%
}

\def\section{\@ifstar{\@ssection}{\@section}}

\def\@section#1{% heading A (\section{....})
  \if@nobreak
    \everypar{}%
    \ifnum\LastMac=\Hae \addvspace{\half}\fi
  \else
    \addpen{\gds@cbrk}%
    \addvspace{\two}%
  \fi
  \bgroup
    \ninepoint\bf
    \Raggedright
    \global\advance\Sec \@ne
    \ifappendix
      \global\Eqnno=\z@ \global\SubEqnno=\z@\relax
      \def\ch@ck{#1}%
      \ifx\ch@ck\empty \def\c@lon{}\else\def\c@lon{:}\fi
      \noindent\@nohdbrk APPENDIX\ \thesection\c@lon\hskip 0.5em%
        \uppercase{#1}\par
    \else
      \noindent\@nohdbrk\thesection\hskip 1pc \uppercase{#1}\par
    \fi
    \global\SecSec=\z@
  \egroup
  \nobreak
  \vskip\half
  \nobreak
  \@noafterindent
  \LastMac=\Hae\relax
}

\def\@ssection#1{%  main section heading (\section*{....})
  \if@nobreak
    \everypar{}%
    \ifnum\LastMac=\Hae \addvspace{\half}\fi
  \else
    \addpen{\gds@cbrk}%
    \addvspace{\two}%
  \fi
  \bgroup
    \ninepoint\bf
    \Raggedright
%    \ifappendix
%      \global\Eqnno=\z@ \global\SubEqnno=\z@\relax % mh in apps dont reset
%      \noindent\@nohdbrk APPENDIX:\hskip 0.5em%
%        \uppercase{#1}\par
%    \else
    \noindent\@nohdbrk\uppercase{#1}\par
%    \fi
  \egroup
  \nobreak
  \vskip\half
  \nobreak
  \@noafterindent
  \LastMac=\Hae\relax
}

\def\subsection{\@ifstar{\@ssubsection}{\@subsection}}

\def\@subsection#1{% heading B
  \if@nobreak
    \everypar{}%
    \ifnum\LastMac=\Hae \addvspace{1pt plus 1pt minus .5pt}\fi
  \else
    \addpen{\gds@cbrk}%
    \addvspace{\onehalf}%
  \fi
  \bgroup
    \ninepoint\bf
    \Raggedright
    \global\advance\SecSec \@ne
    \noindent\@nohdbrk\thesubsection \hskip 1pc\relax #1\par
    \global\SecSecSec=\z@
  \egroup
  \nobreak
  \vskip\half
  \nobreak
  \@noafterindent
  \LastMac=\Hbe\relax
}

\def\@ssubsection#1{% heading B*
  \if@nobreak
    \everypar{}%
    \ifnum\LastMac=\Hae \addvspace{1pt plus 1pt minus .5pt}\fi
  \else
    \addpen{\gds@cbrk}%
    \addvspace{\onehalf}%
  \fi
  \bgroup
    \ninepoint\bf
    \Raggedright
    \noindent\@nohdbrk #1\par
  \egroup
  \nobreak
  \vskip\half
  \nobreak
  \@noafterindent
  \LastMac=\Hbe\relax
}

\def\subsubsection{\@ifstar{\@ssubsubsection}{\@subsubsection}}

\def\@subsubsection#1{% heading C
  \if@nobreak
    \everypar{}%
    \ifnum\LastMac=\Hbe \addvspace{1pt plus 1pt minus .5pt}\fi
  \else
    \addpen{\gds@cbrk}%
    \addvspace{\onehalf}%
  \fi
  \bgroup
    \ninepoint\it
    \Raggedright
    \global\advance\SecSecSec \@ne
    \noindent\@nohdbrk\thesubsubsection \hskip 1pc\relax #1\par
  \egroup
  \nobreak
  \vskip\half
  \nobreak
  \@noafterindent
  \LastMac=\Hce\relax
}

\def\@ssubsubsection#1{% heading C*
  \if@nobreak
    \everypar{}%
    \ifnum\LastMac=\Hbe \addvspace{1pt plus 1pt minus .5pt}\fi
  \else
    \addpen{\gds@cbrk}%
    \addvspace{\onehalf}%
  \fi
  \bgroup
    \ninepoint\it
    \Raggedright
    \noindent\@nohdbrk #1\par
  \egroup
  \nobreak
  \vskip\half
  \nobreak
  \@noafterindent
  \LastMac=\Hce\relax
}

\def\paragraph#1{% heading D
  \if@nobreak
    \everypar{}%
  \else
    \addpen{\gds@cbrk}%
    \addvspace{\one}%
  \fi%
  \bgroup%
    \ninepoint\it
    \noindent #1\ \nobreak%
  \egroup
  \LastMac=\Hde\relax
  \ignorespaces
}

% Appendix

\newif\ifappendix

\def\appendix{%
  \global\appendixtrue
  \def\thesection{\Alph{Sec}}%
  \def\thesubsection{\thesection\arabic{SecSec}}%
  \def\theeq{\thesection\arabic{Eqnno}}%
  \Sec=\z@ \SecSec=\z@ \SecSecSec=\z@ \Eqnno=\z@ \SubEqnno=\z@\relax
}

% Text

 % provided for backward compatibility

% Lists

\def\beginlist{%
  \par\if@nobreak \else\addvspace{\half}\fi%
  \bgroup%
    \ninepoint
    \let\item=\list@item%
}

\def\list@item{%
  \par\noindent\hskip 1em\relax%
  \ignorespaces%
}

\def\endlist{\par\egroup\addvspace{\half}\@doendpe}

% References

\def\beginrefs{%
  \par
  \bgroup
    \eightpoint
    \Fullout
    \let\bibitem=\bib@item
}

\def\bib@item{%
  \par\parindent=1.5em\Hang{1.5em}{1}%
  \everypar={\Hang{1.5em}{1}\ignorespaces}%
  \noindent\ignorespaces
}

\def\endrefs{\par\egroup\@doendpe}

% Page heads

\newtoks\CatchLine

\def\@journal{Mon.\ Not.\ R.\ Astron.\ Soc.\ }  % The journal title string
\def\@pubyear{1994}        % Assign a default publication year
\def\@pagerange{000--000}  % Assign a default page-range
\def\@volume{000}          % Assign a default volume number
\def\@microfiche{}         %

\def\pubyear#1{\gdef\@pubyear{#1}\@makecatchline}
\def\pagerange#1{\gdef\@pagerange{#1}\@makecatchline}
\def\volume#1{\gdef\@volume{#1}\@makecatchline}
\def\microfiche#1{\gdef\@microfiche{and Microfiche\ #1}\@makecatchline}

\def\@makecatchline{%
  \global\CatchLine{%
    {\rm \@journal {\bf \@volume},\ \@pagerange\ (\@pubyear)\ \@microfiche}}%
}

\@makecatchline % Assign a catchline, using the above defaults

\newtoks\LeftHeader
\def\shortauthor#1{% left page head
  \global\LeftHeader{#1}%
}

\newtoks\RightHeader
\def\shorttitle#1{% right page head
  \global\RightHeader{#1}%
}

\def\PageHead{% recto/verso running heads
  \begingroup
    \ifsp@page
      \csname ps@\sp@type\endcsname
    \fi
    \ifodd\pageno
      \let\the@head=\@oddhead
    \else
      \let\the@head=\@evenhead
    \fi
    \vbox to \z@{\vskip-22.5\p@%
      \hbox to \PageWidth{\vbox to8.5\p@{}%
        \the@head
      }%
    \vss}%
  \endgroup
  \nointerlineskip
}

\gdef\PageFoot{%
  \nointerlineskip%
  \begingroup
  \ifsp@page
    \csname ps@\sp@type\endcsname
    \global\sp@pagefalse
  \fi
  \vbox to 22pt{\vfil%
    \hbox to \PageWidth{%
      \eightpoint\strut\noindent
      \ifodd\pageno
        \@oddfoot
      \else
        \@evenfoot
      \fi
    }%
  }%
  \endgroup
}

\def\today{%
  \number\day\space
  \ifcase\month\or January\or February\or March\or April\or May\or June\or
    July\or August\or September\or October\or November\or December\fi
  \space\number\year%
}

\def\authorcomment#1{%
  \gdef\PageFoot{%
    \nointerlineskip%
    \vbox to 20pt{\vfil%
      \hbox to \PageWidth{\elevenpoint\noindent \hfil #1 \hfil}}%
  }%
}

% Plate pages

\newif\ifplate@page
\newbox\plt@box

\def\beginplatepage{%
  \let\plate=\plate@head
  \let\caption=\fig@caption
  \global\setbox\plt@box=\vbox\bgroup
  \TEMPDIMEN=\PageWidth % For \fig@caption test
  \hsize=\PageWidth\relax
}

\def\endplatepage{\par\egroup\global\plate@pagetrue}
\def\plate@head#1{\gdef\plt@cap{#1}}

% Letters option

\def\letters{%
  \gdef\folio{\ifnum\pageno<\z@ L\romannumeral-\pageno
    \else L\number\pageno \fi}%
}

% Math setup

% The standard math indentation
\newdimen\mathindent

\global\mathindent=\z@
\global\everydisplay{\global\@dspwd=\displaywidth\displaysetup}

% New versions of \displaylines, \eqalign, \eqalignno for
% when non-centered math is in use.

\def\@displaylines#1{% (for non-centered math)
  {}$\displ@y\hbox{\vbox{\halign{$\@lign\hfil\displaystyle##\hfil$\crcr
  #1\crcr}}}${}%
}

\def\@eqalign#1{\null\vcenter{\openup\jot\m@th% (for non-centered math)
  \ialign{\strut\hfil$\displaystyle{##}$&$\displaystyle{{}##}$\hfil
      \crcr#1\crcr}}%
}

\def\@eqalignno#1{% (for non-centered math)
  \global\advance\@dspwd by -\mathindent%
  {}$\displ@y\hbox{\vbox{\halign to\@dspwd%
  {\hfil$\@lign\displaystyle{##}$\tabskip\z@skip
  &$\@lign\displaystyle{{}##}$\hfil\tabskip\centering
  &\llap{$\@lign##$}\tabskip\z@skip\crcr
  #1\crcr}}}${}%
}

% When equations are flushleft ensure, that \displaylines,
% \eqalign, \eqalignno and \leqalignno point to the new versions of
% the macros. Also make \leqalignno act like \eqalignno, otherwise the
% equation text would `crash' into the equation number.

\global\let\displaylines=\@displaylines
\global\let\eqalign=\@eqalign
\global\let\eqalignno=\@eqalignno
\global\let\leqalignno=\@eqalignno

\newdimen\@dspwd   \@dspwd=\z@
\newif\if@eqno
\newif\if@leqno
\newtoks\@eqn
\newtoks\@eq

\def\displaysetup#1$${\displaytest#1\eqno\eqno\displaytest}

\def\displaytest#1\eqno#2\eqno#3\displaytest{%
 \if!#3!\ldisplaytest#1\leqno\leqno\ldisplaytest
 \else\@eqnotrue\@leqnofalse\@eqn={#2}\@eq={#1}\fi
 \generaldisplay$$}

\def\ldisplaytest#1\leqno#2\leqno#3\ldisplaytest{%
\@eq={#1}%
 \if!#3!\@eqnofalse\else\@eqnotrue\@leqnotrue
  \@eqn={#2}\fi}

\def\generaldisplay{%
  \if@eqno
    \if@leqno
      \hbox to \displaywidth{\noindent
        \rlap{$\displaystyle\the\@eqn$}%
        \hskip\mathindent$\displaystyle\the\@eq$\hfil}%
    \else
      \hbox to \displaywidth{\noindent
        \hskip\mathindent
        $\displaystyle\the\@eq$\hfil$\displaystyle\the\@eqn$}%
    \fi
  \else
    \hbox to \displaywidth{\noindent
      \hskip\mathindent$\displaystyle\the\@eq$\hfil}%
  \fi
}

% Finishing notice

\def\@notice{%
  \par\Two%
  \noindent{\b@ls{11pt}\ninerm This paper has been produced using the
    Royal Astronomical Society/Blackwell Science \TeX\ macros.\par}%
}

% redefine \bye to output our identification notice :
\outer\def\bye{\@notice\par\vfill\supereject\end}

% define a sign on :

\def\start@mess{%
  Monthly notices of the RAS journal style (\@typeface)\space
    v\@version,\space \@verdate.%
}

\everyjob{\Warn{\start@mess}}

% Two-column macros

%--------------------------------------------------------%
%                     INITIALISATION                     %
%--------------------------------------------------------%

\newif\if@debug \@debugfalse  %  when false, only warnings displayed

\def\Print#1{\if@debug\immediate\write16{#1}\else \fi}
\def\Warn#1{\immediate\write16{#1}}
\def\wlog#1{}

\newcount\Iteration % temporary loop counter

\def\Single{0} \def\Double{1}                 % ItemSPAN's
\def\Figure{0} \def\Table{1}                  % ItemTYPE's

\def\InStack{0}  % ItemSTATUS
\def\InZoneA{1}
\def\InZoneB{2}
\def\InZoneC{3}

\newcount\TEMPCOUNT % temporary count register
\newdimen\TEMPDIMEN % temporary dimen register
\newbox\TEMPBOX     % temporary box register
\newbox\VOIDBOX     % a box which is permenately void

\newcount\LengthOfStack % number of items currently in stack
\newcount\MaxItems      % maximum number of items allowed in stack
\newcount\StackPointer
\newcount\Point         % used in calculation for generating the
                        % physical address of an item in the stack.
\newcount\NextFigure    % number of next figure to be output
\newcount\NextTable     % number of next table to be output
\newcount\NextItem      % Next item to consider by order in stack

\newcount\StatusStack   % set to point to top of STATUS stack
\newcount\NumStack      % set to point to top of NUMBER stack
\newcount\TypeStack     % set to point to top of TYPE stack
\newcount\SpanStack     % set to point to top of SPAN stack
\newcount\BoxStack      % set to point to top of BOX stack

\newcount\ItemSTATUS    % status of present item
\newcount\ItemNUMBER    % number of present item
\newcount\ItemTYPE      % type of present item
\newcount\ItemSPAN      % span of present item
\newbox\ItemBOX         % box of present item
\newdimen\ItemSIZE      % size of present item
                        % (calculated by GetItemBOX)

\newdimen\PageHeight    % vertical measure of full page
\newdimen\TextLeading   % distance between baselines of body text
\newdimen\Feathering    % amount of interline stretch
\newcount\LinesPerPage  % height of page in text lines
\newdimen\ColumnWidth   % width of 1 column of text
\newdimen\ColumnGap     % size of gap between columns
\newdimen\PageWidth     % = \ColumnWidth * 2 + \ColumnGap
\newdimen\BodgeHeight   % Bodge to align figures and tables with text
\newcount\Leading       % Set to same as \TextLeading above

\newdimen\ZoneBSize  % size of items in ZoneB
\newdimen\TextSize   % size of text in ZoneB
\newbox\ZoneABOX     % contains Zone A material
\newbox\ZoneBBOX     % contains Zone B material
\newbox\ZoneCBOX     % contains Zone C material

\newif\ifFirstSingleItem
\newif\ifFirstZoneA
\newif\ifMakePageInComplete
\newif\ifMoreFigures \MoreFiguresfalse % set true in join stack
\newif\ifMoreTables  \MoreTablesfalse  % set true in join stack

\newif\ifFigInZoneB % used to determine in which zone an item
\newif\ifFigInZoneC % will be placed based on what is in other
\newif\ifTabInZoneB % zones already for a given page.
\newif\ifTabInZoneC

\newif\ifZoneAFullPage

\newbox\MidBOX    % = LeftBOX+gap+RightBOX
\newbox\LeftBOX
\newbox\RightBOX
\newbox\PageBOX   % complete made-up page

\newif\ifLeftCOL  % flags first pass through output routine
\LeftCOLtrue

\newdimen\ZoneBAdjust

\newcount\ItemFits
\def\Yes{1}
\def\No{2}

% Setup file.

\MaxItems=15
\NextFigure=\z@        % used for article opening
\NextTable=\@ne

\BodgeHeight=6pt
\TextLeading=11pt    % baselineskip of body text
\Leading=11
\Feathering=\z@      % amount of interline stretch
\LinesPerPage=61     % number of text lines per full page -1
\topskip=\TextLeading
\ColumnWidth=20pc    % width of text columns
\ColumnGap=2pc       % gap between columns

\newskip\ItemSepamount  % space between floats
\ItemSepamount=\TextLeading plus \TextLeading minus 4pt

\parskip=\z@ plus .1pt
\parindent=18pt
\widowpenalty=\z@
\clubpenalty=10000
\tolerance=1500
\hbadness=1500
\abovedisplayskip=6pt plus 2pt minus 1pt
\belowdisplayskip=6pt plus 2pt minus 1pt
\abovedisplayshortskip=6pt plus 2pt minus 1pt
\belowdisplayshortskip=6pt plus 2pt minus 1pt

\frenchspacing

\ninepoint % start main text size

\PageHeight=682pt
\PageWidth=2\ColumnWidth
\advance\PageWidth by \ColumnGap

\pagestyle{headings}

%--------------------------------------------------------%
%                         STACKS                         %
%--------------------------------------------------------%

% THE ITEM STACK
% The item stack contains contains figures and tables
% in the order in which they appear in the article source
% code.

% allocate stack space

\newcount\DUMMY \StatusStack=\allocationnumber
\newcount\DUMMY \newcount\DUMMY \newcount\DUMMY 
\newcount\DUMMY \newcount\DUMMY \newcount\DUMMY 
\newcount\DUMMY \newcount\DUMMY \newcount\DUMMY
\newcount\DUMMY \newcount\DUMMY \newcount\DUMMY 
\newcount\DUMMY \newcount\DUMMY \newcount\DUMMY

\newcount\DUMMY \NumStack=\allocationnumber
\newcount\DUMMY \newcount\DUMMY \newcount\DUMMY 
\newcount\DUMMY \newcount\DUMMY \newcount\DUMMY 
\newcount\DUMMY \newcount\DUMMY \newcount\DUMMY 
\newcount\DUMMY \newcount\DUMMY \newcount\DUMMY 
\newcount\DUMMY \newcount\DUMMY \newcount\DUMMY

\newcount\DUMMY \TypeStack=\allocationnumber
\newcount\DUMMY \newcount\DUMMY \newcount\DUMMY 
\newcount\DUMMY \newcount\DUMMY \newcount\DUMMY 
\newcount\DUMMY \newcount\DUMMY \newcount\DUMMY 
\newcount\DUMMY \newcount\DUMMY \newcount\DUMMY 
\newcount\DUMMY \newcount\DUMMY \newcount\DUMMY

\newcount\DUMMY \SpanStack=\allocationnumber
\newcount\DUMMY \newcount\DUMMY \newcount\DUMMY 
\newcount\DUMMY \newcount\DUMMY \newcount\DUMMY 
\newcount\DUMMY \newcount\DUMMY \newcount\DUMMY 
\newcount\DUMMY \newcount\DUMMY \newcount\DUMMY 
\newcount\DUMMY \newcount\DUMMY \newcount\DUMMY

\newbox\DUMMY   \BoxStack=\allocationnumber
\newbox\DUMMY   \newbox\DUMMY \newbox\DUMMY 
\newbox\DUMMY   \newbox\DUMMY \newbox\DUMMY 
\newbox\DUMMY   \newbox\DUMMY \newbox\DUMMY 
\newbox\DUMMY   \newbox\DUMMY \newbox\DUMMY 
\newbox\DUMMY   \newbox\DUMMY \newbox\DUMMY

\def\wlog{\immediate\write\m@ne}

% \GetItemSTATUS, \GetItemNUMBER, \GetItemTYPE, \GetItemSPAN,
% \GetItemBox 
% are used to get details of a particular item from the item
% stack. The argument to each of these is the items position
% in the stack (usually \StackPointer)...not the items number.

\def\GetItemAll#1{%
 \GetItemSTATUS{#1}
 \GetItemNUMBER{#1}
 \GetItemTYPE{#1}
 \GetItemSPAN{#1}
 \GetItemBOX{#1}
}

% Note: \LeaveStack uses this routine. Do not destroy \Point
\def\GetItemSTATUS#1{%
 \Point=\StatusStack
 \advance\Point by #1
 \global\ItemSTATUS=\count\Point
}

% Note: \LeaveStack uses this routine. Do not destroy \Point
\def\GetItemNUMBER#1{%
 \Point=\NumStack
 \advance\Point by #1
 \global\ItemNUMBER=\count\Point
}

% Note: \LeaveStack uses this routine. Do not destroy \Point
\def\GetItemTYPE#1{%
 \Point=\TypeStack
 \advance\Point by #1
 \global\ItemTYPE=\count\Point
}

% Note: \LeaveStack uses this routine. Do not destroy \Point
\def\GetItemSPAN#1{%
 \Point\SpanStack
 \advance\Point by #1
 \global\ItemSPAN=\count\Point
}

% Note: \LeaveStack uses this routine. Do not destroy \Point
\def\GetItemBOX#1{%
 \Point=\BoxStack
 \advance\Point by #1
 \global\setbox\ItemBOX=\vbox{\copy\Point}
 \global\ItemSIZE=\ht\ItemBOX
 \global\advance\ItemSIZE by \dp\ItemBOX
 \TEMPCOUNT=\ItemSIZE
 \divide\TEMPCOUNT by \Leading
 \divide\TEMPCOUNT by 65536
 \advance\TEMPCOUNT \@ne
 \ItemSIZE=\TEMPCOUNT pt
 \global\multiply\ItemSIZE by \Leading
}

% item joins stack

\def\JoinStack{%
 \ifnum\LengthOfStack=\MaxItems % stack is full of items
  \Warn{WARNING: Stack is full...some items will be lost!}
 \else
  \Point=\StatusStack
  \advance\Point by \LengthOfStack
  \global\count\Point=\ItemSTATUS
  \Point=\NumStack
  \advance\Point by \LengthOfStack
  \global\count\Point=\ItemNUMBER
  \Point=\TypeStack
  \advance\Point by \LengthOfStack
  \global\count\Point=\ItemTYPE
  \Point\SpanStack
  \advance\Point by \LengthOfStack
  \global\count\Point=\ItemSPAN
  \Point=\BoxStack
  \advance\Point by \LengthOfStack
  \global\setbox\Point=\vbox{\copy\ItemBOX}
  \global\advance\LengthOfStack \@ne
  \ifnum\ItemTYPE=\Figure % used in \MakePage
   \global\MoreFigurestrue
  \else
   \global\MoreTablestrue
  \fi
 \fi
}

% item leaves stack
% #1=physical position of the item to be removed

\def\LeaveStack#1{%
 {\Iteration=#1
 \loop
 \ifnum\Iteration<\LengthOfStack
  \advance\Iteration \@ne
  \GetItemSTATUS{\Iteration}
   \advance\Point by \m@ne
   \global\count\Point=\ItemSTATUS
  \GetItemNUMBER{\Iteration}
   \advance\Point by \m@ne
   \global\count\Point=\ItemNUMBER
  \GetItemTYPE{\Iteration}
   \advance\Point by \m@ne
   \global\count\Point=\ItemTYPE
  \GetItemSPAN{\Iteration}
   \advance\Point by \m@ne
   \global\count\Point=\ItemSPAN
  \GetItemBOX{\Iteration}
   \advance\Point by \m@ne
   \global\setbox\Point=\vbox{\copy\ItemBOX}
 \repeat}
 \global\advance\LengthOfStack by \m@ne
}

% clean stack
% This routine scans through the stack and removes anything
% that does not have STATUS=\InStack.

\newif\ifStackNotClean

\def\CleanStack{%
 \StackNotCleantrue
 {\Iteration=\z@
  \loop
   \ifStackNotClean
    \GetItemSTATUS{\Iteration}
    \ifnum\ItemSTATUS=\InStack
     \advance\Iteration \@ne
     \else
      \LeaveStack{\Iteration}
    \fi
   \ifnum\LengthOfStack<\Iteration
    \StackNotCleanfalse
   \fi
 \repeat}
}

% Find item.
% This macro searches from the top to the bottom of the
% stack for an item of a specified type and number.
% #1=type, #2=number
% If the specified item is found, then \StackPointer is set
% to point to it, else \StackPointer=-1.
% This routine is used to find the next figure or table
% by number.

\def\FindItem#1#2{%
 \global\StackPointer=\m@ne % assume item isn't in stack for now
 {\Iteration=\z@
  \loop
  \ifnum\Iteration<\LengthOfStack
   \GetItemSTATUS{\Iteration}
   \ifnum\ItemSTATUS=\InStack
    \GetItemTYPE{\Iteration}
    \ifnum\ItemTYPE=#1
     \GetItemNUMBER{\Iteration}
     \ifnum\ItemNUMBER=#2
      \global\StackPointer=\Iteration
      \Iteration=\LengthOfStack % terminate loop
     \fi
    \fi
   \fi
  \advance\Iteration \@ne
 \repeat}
}

% Find next type
% This macro searches from the top to the bottom of the stack
% looking for the first item which has STATUS=\InStack.
% If it is a figure then a figure is what will be considered
% next by \MakePage else table.

\def\FindNext{%
 \global\StackPointer=\m@ne % assume stack is empty for now
 {\Iteration=\z@
  \loop
  \ifnum\Iteration<\LengthOfStack
   \GetItemSTATUS{\Iteration}
   \ifnum\ItemSTATUS=\InStack
    \GetItemTYPE{\Iteration}
   \ifnum\ItemTYPE=\Figure
    \ifMoreFigures
      \global\NextItem=\Figure
      \global\StackPointer=\Iteration
      \Iteration=\LengthOfStack % terminate loop
    \fi
   \fi
   \ifnum\ItemTYPE=\Table
    \ifMoreTables
      \global\NextItem=\Table
      \global\StackPointer=\Iteration
      \Iteration=\LengthOfStack % terminate loop
    \fi
   \fi
  \fi
  \advance\Iteration \@ne
 \repeat}
}

% Change status
% Macro to change the status of a specified item in stack.
% #1=item, #2=new status

\def\ChangeStatus#1#2{%
 \Point=\StatusStack
 \advance\Point by #1
 \global\count\Point=#2
}

%--------------------------------------------------------%
%                       MAKEPAGE                         %
%--------------------------------------------------------%

% This macro is called at the start of every new page
% including the first. It scans through the stack picking
% out items which should be placed on this page. It then
% leaves space for the items to be placed later. The routine
% terminates when either there is no room on the page to
% fit the next figure or table, or there are no more items
% in the stack.

\def\Zone{\InZoneA}

\ZoneBAdjust=\z@

\def\MakePage{% allocate space on this page for stack items
 \global\ZoneBSize=\PageHeight
 \global\TextSize=\ZoneBSize
 \global\ZoneAFullPagefalse
 \global\topskip=\TextLeading
 \MakePageInCompletetrue
 \MoreFigurestrue
 \MoreTablestrue
 \FigInZoneBfalse
 \FigInZoneCfalse
 \TabInZoneBfalse
 \TabInZoneCfalse
 \global\FirstSingleItemtrue
 \global\FirstZoneAtrue
 \global\setbox\ZoneABOX=\box\VOIDBOX
 \global\setbox\ZoneBBOX=\box\VOIDBOX
 \global\setbox\ZoneCBOX=\box\VOIDBOX
 \loop
  \ifMakePageInComplete
 \FindNext
 \ifnum\StackPointer=\m@ne
  \NextItem=\m@ne
  \MoreFiguresfalse
  \MoreTablesfalse
 \fi
 \ifnum\NextItem=\Figure
   \FindItem{\Figure}{\NextFigure}
   \ifnum\StackPointer=\m@ne \global\MoreFiguresfalse
   \else
    \GetItemSPAN{\StackPointer}
    \ifnum\ItemSPAN=\Single \def\Zone{\InZoneB}\relax
     \ifFigInZoneC \global\MoreFiguresfalse\fi
    \else
     \def\Zone{\InZoneA}
     \ifFigInZoneB \def\Zone{\InZoneC}\fi
    \fi
   \fi
   \ifMoreFigures\Print{}\FigureItems\fi
 \fi
\ifnum\NextItem=\Table
   \FindItem{\Table}{\NextTable}
   \ifnum\StackPointer=\m@ne \global\MoreTablesfalse
   \else
    \GetItemSPAN{\StackPointer}
    \ifnum\ItemSPAN=\Single\relax
     \ifTabInZoneC \global\MoreTablesfalse\fi
    \else
     \def\Zone{\InZoneA}
     \ifTabInZoneB \def\Zone{\InZoneC}\fi
    \fi
   \fi
   \ifMoreTables\Print{}\TableItems\fi
 \fi
   \MakePageInCompletefalse % assume page is complete
   \ifMoreFigures\MakePageInCompletetrue\fi
   \ifMoreTables\MakePageInCompletetrue\fi
 \repeat
%\Print{TextSize=\the\TextSize}
%\Print{ZoneBSize=\the\ZoneBSize}
 \ifZoneAFullPage
  \global\TextSize=\z@
  \global\ZoneBSize=\z@
  \global\vsize=\z@\relax
  \global\topskip=\z@\relax
  \vbox to \z@{\vss}
  \eject
 \else
 \global\advance\ZoneBSize by -\ZoneBAdjust
 \global\vsize=\ZoneBSize
 \global\hsize=\ColumnWidth
 \global\ZoneBAdjust=\z@
 \ifdim\TextSize<23pt
 \Warn{}
 \Warn{* Making column fall short: TextSize=\the\TextSize *}
 \vskip-\lastskip\eject\fi
 \fi
}

\def\MakeRightCol{% allocate space for the right column of text
 \global\TextSize=\ZoneBSize
 \MakePageInCompletetrue
 \MoreFigurestrue
 \MoreTablestrue
 \global\FirstSingleItemtrue
 \global\setbox\ZoneBBOX=\box\VOIDBOX
 \def\Zone{\InZoneB}
 \loop
  \ifMakePageInComplete
 \FindNext
 \ifnum\StackPointer=\m@ne
  \NextItem=\m@ne
  \MoreFiguresfalse
  \MoreTablesfalse
 \fi
 \ifnum\NextItem=\Figure
   \FindItem{\Figure}{\NextFigure}
   \ifnum\StackPointer=\m@ne \MoreFiguresfalse
   \else
    \GetItemSPAN{\StackPointer}
    \ifnum\ItemSPAN=\Double\relax
     \MoreFiguresfalse\fi
   \fi
   \ifMoreFigures\Print{}\FigureItems\fi
 \fi
 \ifnum\NextItem=\Table
   \FindItem{\Table}{\NextTable}
   \ifnum\StackPointer=\m@ne \MoreTablesfalse
   \else
    \GetItemSPAN{\StackPointer}
    \ifnum\ItemSPAN=\Double\relax
     \MoreTablesfalse\fi
   \fi
   \ifMoreTables\Print{}\TableItems\fi
 \fi
   \MakePageInCompletefalse % assume page is complete
   \ifMoreFigures\MakePageInCompletetrue\fi
   \ifMoreTables\MakePageInCompletetrue\fi
 \repeat
 \ifZoneAFullPage
  \global\TextSize=\z@
  \global\ZoneBSize=\z@
  \global\vsize=\z@\relax
  \global\topskip=\z@\relax
  \vbox to \z@{\vss}
  \eject
 \else
 \global\vsize=\ZoneBSize
 \global\hsize=\ColumnWidth
 \ifdim\TextSize<23pt
 \Warn{}
 \Warn{* Making column fall short: TextSize=\the\TextSize *}
 \vskip-\lastskip\eject\fi
\fi
}

\def\FigureItems{% Stack pointer points to next figure
 \Print{Considering...}
 \ShowItem{\StackPointer}
 \GetItemBOX{\StackPointer} % auto calculates ItemSIZE
 \GetItemSPAN{\StackPointer}
  \CheckFitInZone % check to see if item fits
  \ifnum\ItemFits=\Yes
   \ifnum\ItemSPAN=\Single
     \ChangeStatus{\StackPointer}{\InZoneB} % flag to be output
     \global\FigInZoneBtrue
     \ifFirstSingleItem
      \hbox{}\vskip-\BodgeHeight
     \global\advance\ItemSIZE by \TextLeading
     \fi
     \unvbox\ItemBOX\ItemSep
     \global\FirstSingleItemfalse
     \global\advance\TextSize by -\ItemSIZE% allocate space
     \global\advance\TextSize by -\TextLeading
   \else
    \ifFirstZoneA
     \global\advance\ItemSIZE by \TextLeading
     \global\FirstZoneAfalse\fi
    \global\advance\TextSize by -\ItemSIZE
    \global\advance\TextSize by -\TextLeading
    \global\advance\ZoneBSize by -\ItemSIZE
    \global\advance\ZoneBSize by -\TextLeading
    \ifFigInZoneB\relax
     \else
     \ifdim\TextSize<3\TextLeading
     \global\ZoneAFullPagetrue
     \fi
    \fi
    \ChangeStatus{\StackPointer}{\Zone}
    \ifnum\Zone=\InZoneC \global\FigInZoneCtrue\fi
  \fi
   \Print{TextSize=\the\TextSize}
   \Print{ZoneBSize=\the\ZoneBSize}
  \global\advance\NextFigure \@ne
   \Print{This figure has been placed.}
  \else
   \Print{No space available for this figure...holding over.}
   \Print{}
   \global\MoreFiguresfalse
  \fi
}

\def\TableItems{% Stack pointer points to next table
 \Print{Considering...}
 \ShowItem{\StackPointer}
 \GetItemBOX{\StackPointer} % auto calculates ItemSIZE
 \GetItemSPAN{\StackPointer}
  \CheckFitInZone % check to see of item fits in Zone
  \ifnum\ItemFits=\Yes
   \ifnum\ItemSPAN=\Single
    \ChangeStatus{\StackPointer}{\InZoneB}
     \global\TabInZoneBtrue
     \ifFirstSingleItem
      \hbox{}\vskip-\BodgeHeight
     \global\advance\ItemSIZE by \TextLeading
     \fi
     \unvbox\ItemBOX\ItemSep
     \global\FirstSingleItemfalse
     \global\advance\TextSize by -\ItemSIZE
     \global\advance\TextSize by -\TextLeading
   \else
    \ifFirstZoneA
    \global\advance\ItemSIZE by \TextLeading
    \global\FirstZoneAfalse\fi
    \global\advance\TextSize by -\ItemSIZE
    \global\advance\TextSize by -\TextLeading
    \global\advance\ZoneBSize by -\ItemSIZE
    \global\advance\ZoneBSize by -\TextLeading
    \ifFigInZoneB\relax
     \else
     \ifdim\TextSize<3\TextLeading
     \global\ZoneAFullPagetrue
     \fi
    \fi
    \ChangeStatus{\StackPointer}{\Zone}
    \ifnum\Zone=\InZoneC \global\TabInZoneCtrue\fi
   \fi
%   \Print{TextSize=\the\TextSize}
%   \Print{ZoneBSize=\the\ZoneBSize}
  \global\advance\NextTable \@ne
   \Print{This table has been placed.}
  \else
  \Print{No space available for this table...holding over.}
   \Print{}
   \global\MoreTablesfalse
  \fi
}

% These macros check to see if an item of ItemSIZE will
% fit in a particular zone. If it will, then ItemFits
% will be set true else false.

\def\CheckFitInZone{%
{\advance\TextSize by -\ItemSIZE
 \advance\TextSize by -\TextLeading
 \ifFirstSingleItem
  \advance\TextSize by \TextLeading
 \fi
 \ifnum\Zone=\InZoneA\relax
  \else \advance\TextSize by -\ZoneBAdjust
 \fi
 \ifdim\TextSize<3\TextLeading \global\ItemFits=\No
 \else \global\ItemFits=\Yes\fi}
}

\def\BeginOpening{%
  % start 9pt a.s.a.p. so that \New.. commands get a chance to init.
  \ninepoint
  \thispagestyle{titlepage}%
  \global\setbox\ItemBOX=\vbox\bgroup%
    \hsize=\PageWidth%
    \hrule height \z@
    \ifsinglecol\vskip 6pt\fi % Bodge, to get same pos. as two-column!
}

\let\begintopmatter=\BeginOpening  %  alias for \BeginOpening

\def\EndOpening{%
  \One%  1 line fixed space below opening
  \egroup
  \ifsinglecol
    \box\ItemBOX%
    \vskip\TextLeading plus 2\TextLeading% var. space: min 1, max 3 lines
    \@noafterindent
  \else
    \ItemNUMBER=\z@%
    \ItemTYPE=\Figure
    \ItemSPAN=\Double
    \ItemSTATUS=\InStack
    \JoinStack
  \fi
}

% Figures

\newif\if@here  \@herefalse

\def\no@float{\global\@heretrue}
\let\nofloat=\relax % only enabled for one column

\def\beginfigure{%
  \@ifstar{\global\@dfloattrue \@bfigure}{\global\@dfloatfalse \@bfigure}%
}

\def\@bfigure#1{%
  \par
  \if@dfloat
    \ItemSPAN=\Double
    \TEMPDIMEN=\PageWidth
  \else
    \ItemSPAN=\Single
    \TEMPDIMEN=\ColumnWidth
  \fi
  \ifsinglecol
    \TEMPDIMEN=\PageWidth
  \else
    \ItemSTATUS=\InStack
    \ItemNUMBER=#1%
    \ItemTYPE=\Figure
  \fi
  \bgroup
    \hsize=\TEMPDIMEN
    \global\setbox\ItemBOX=\vbox\bgroup
      \eightpoint\nostb@ls{10pt}%
      \let\caption=\fig@caption
      \ifsinglecol \let\nofloat=\no@float\fi
}

\def\fig@caption#1{%
  \vskip 5.5pt plus 6pt%
  \bgroup % grouping and size change needed for plate pages
    \eightpoint\nostb@ls{10pt}%
    \setbox\TEMPBOX=\hbox{#1}%
    \ifdim\wd\TEMPBOX>\TEMPDIMEN
      \noindent \unhbox\TEMPBOX\par
    \else
      \hbox to \hsize{\hfil\unhbox\TEMPBOX\hfil}%
    \fi
  \egroup
}

\def\endfigure{%
  \par\egroup % end \vbox
  \egroup
  \ifsinglecol
    \if@here \midinsert\global\@herefalse\else \topinsert\fi
      \unvbox\ItemBOX
    \endinsert
  \else
    \JoinStack
    \Print{Processing source for figure \the\ItemNUMBER}%
  \fi
}

% Tables

\newbox\tab@cap@box
\def\tab@caption#1{\global\setbox\tab@cap@box=\hbox{#1\par}}

\newtoks\tab@txt@toks
\long\def\tab@txt#1{\global\tab@txt@toks={#1}\global\table@txttrue}

\newif\iftable@txt  \table@txtfalse
\newif\if@dfloat    \@dfloatfalse

\def\begintable{%
  \@ifstar{\global\@dfloattrue \@btable}{\global\@dfloatfalse \@btable}%
}

\def\@btable#1{%
  \par
  \if@dfloat
    \ItemSPAN=\Double
    \TEMPDIMEN=\PageWidth
  \else
    \ItemSPAN=\Single
    \TEMPDIMEN=\ColumnWidth
  \fi
  \ifsinglecol
    \TEMPDIMEN=\PageWidth
  \else
    \ItemSTATUS=\InStack
    \ItemNUMBER=#1%
    \ItemTYPE=\Table
  \fi
  \bgroup
    \eightpoint\nostb@ls{10pt}%
    \global\setbox\ItemBOX=\vbox\bgroup
      \let\caption=\tab@caption
      \let\tabletext=\tab@txt
      \ifsinglecol \let\nofloat=\no@float\fi
}

\def\endtable{%
  \par\egroup % end \vbox
  \egroup
  \setbox\TEMPBOX=\hbox to \TEMPDIMEN{%
    \eightpoint\nostb@ls{10pt}%
    \hss
    \vbox{%
      \hsize=\wd\ItemBOX
      \ifvoid\tab@cap@box
      \else
        \noindent\unhbox\tab@cap@box
        \vskip 5.5pt plus 6pt%
      \fi
      \box\ItemBOX
      \iftable@txt
        \vskip 10pt%
        \noindent\the\tab@txt@toks
        \global\table@txtfalse
      \fi
    }%
    \hss
  }%
  \ifsinglecol
    \if@here \midinsert\global\@herefalse\else \topinsert\fi
      \box\TEMPBOX
    \endinsert
  \else
    \global\setbox\ItemBOX=\box\TEMPBOX
    \JoinStack
    \Print{Processing source for table \the\ItemNUMBER}%
  \fi
}

\def\UnloadZoneA{%
\FirstZoneAtrue
 \Iteration=\z@
  \loop
   \ifnum\Iteration<\LengthOfStack
    \GetItemSTATUS{\Iteration}
    \ifnum\ItemSTATUS=\InZoneA
     \GetItemBOX{\Iteration}
     \ifFirstZoneA \vbox to \BodgeHeight{\vfil}%
     \FirstZoneAfalse\fi
     \unvbox\ItemBOX\ItemSep
     \LeaveStack{\Iteration}
     \else
     \advance\Iteration \@ne
   \fi
 \repeat
}

\def\UnloadZoneC{%
\Iteration=\z@
  \loop
   \ifnum\Iteration<\LengthOfStack
    \GetItemSTATUS{\Iteration}
    \ifnum\ItemSTATUS=\InZoneC
     \GetItemBOX{\Iteration}
     \ItemSep\unvbox\ItemBOX
     \LeaveStack{\Iteration}
     \else
     \advance\Iteration \@ne
   \fi
 \repeat
}

%--------------------------------------------------------%
%                         DIAGNOSTICS                    %
%--------------------------------------------------------%

\def\ShowItem#1{% Show details of on item entry in stack
  {\GetItemAll{#1}
  \Print{\the#1:
  {TYPE=\ifnum\ItemTYPE=\Figure Figure\else Table\fi}
  {NUMBER=\the\ItemNUMBER}
  {SPAN=\ifnum\ItemSPAN=\Single Single\else Double\fi}
  {SIZE=\the\ItemSIZE}}}
}

\def\ShowStack{% 
 \Print{}
 \Print{LengthOfStack = \the\LengthOfStack}
 \ifnum\LengthOfStack=\z@ \Print{Stack is empty}\fi
 \Iteration=\z@
 \loop
 \ifnum\Iteration<\LengthOfStack
  \ShowItem{\Iteration}
  \advance\Iteration \@ne
 \repeat
}

\def\B#1#2{%
\hbox{\vrule\kern-0.4pt\vbox to #2{%
\hrule width #1\vfill\hrule}\kern-0.4pt\vrule}
}

%-------------------------------------------------------%
%             SINGLE COLUMN OUTPUT ROUTINE              %
%-------------------------------------------------------%

\newif\ifsinglecol   \singlecolfalse

\def\onecolumn{%
  \global\output={\singlecoloutput}%
  \global\hsize=\PageWidth
  \global\vsize=\PageHeight
  \global\ColumnWidth=\hsize
  \global\TextLeading=12pt
  \global\Leading=12
  \global\singlecoltrue
  \global\let\onecolumn=\relax%         stop them using \onecolumn again
  \global\let\footnote=\sing@footnote%  enable footnotes
  \global\let\vfootnote=\sing@vfootnote
  \ninepoint % reset \baselineskip after leading change
  \message{(Single column)}%
}

\def\singlecoloutput{%
  \shipout\vbox{\PageHead\vbox to \PageHeight{\pagebody\vss}\PageFoot}%
  \advancepageno
  \ifplate@page
    \shipout\vbox{%
      \sp@pagetrue
      \def\sp@type{plate}%
      \global\plate@pagefalse
      \PageHead\vbox to \PageHeight{\unvbox\plt@box\vfil}\PageFoot%
    }%
    \message{[plate]}%
    \advancepageno
  \fi
  \ifnum\outputpenalty>-\@MM \else\dosupereject\fi%
}

\def\ItemSep{\vskip\ItemSepamount\relax}

\def\ItemSepbreak{\par\ifdim\lastskip<\ItemSepamount
  \removelastskip\penalty-200\ItemSep\fi%
}

% Modify plain's \endinsert so that the mn's spacing is used

\let\@@endinsert=\endinsert % save plain's original \endinsert

\def\endinsert{\egroup % finish the \vbox
  \if@mid \dimen@\ht\z@ \advance\dimen@\dp\z@ \advance\dimen@12\p@
    \advance\dimen@\pagetotal \advance\dimen@-\pageshrink
    \ifdim\dimen@>\pagegoal\@midfalse\p@gefalse\fi\fi
  \if@mid \ItemSep\box\z@\ItemSepbreak
  \else\insert\topins{\penalty100 % floating insertion
    \splittopskip\z@skip
    \splitmaxdepth\maxdimen \floatingpenalty\z@
    \ifp@ge \dimen@\dp\z@
    \vbox to\vsize{\unvbox\z@\kern-\dimen@}% depth is zero
    \else \box\z@\nobreak\ItemSep\fi}\fi\endgroup%
}

% Footnotes (only enabled in single column)

\def\gobbleone#1{}
\def\gobbletwo#1#2{}
\let\footnote=\gobbletwo % Gobble footnote's unless enabled by \onecolumn
\let\vfootnote=\gobbleone

\def\sing@footnote#1{\let\@sf\empty % parameter #2 (the text) is read later
  \ifhmode\edef\@sf{\spacefactor\the\spacefactor}\/\fi
  \hbox{$^{\hbox{\eightpoint #1}}$}\@sf\sing@vfootnote{#1}%
}

\def\sing@vfootnote#1{\insert\footins\bgroup\eightpoint\b@ls{9pt}%
  \interlinepenalty\interfootnotelinepenalty
  \splittopskip\ht\strutbox % top baseline for broken footnotes
  \splitmaxdepth\dp\strutbox \floatingpenalty\@MM
  \leftskip\z@skip \rightskip\z@skip \spaceskip\z@skip \xspaceskip\z@skip
  \noindent $^{\scriptstyle\hbox{#1}}$\hskip 4pt%
    \footstrut\futurelet\next\fo@t%
}

% Kill footnote rule
\def\footnoterule{\kern-3\p@ \hrule height \z@ \kern 3\p@}

\skip\footins=19.5pt plus 12pt minus 1pt
\count\footins=1000
\dimen\footins=\maxdimen

% for footnotes in double column: use \note{$\star$}{footnote}
\def\note#1#2{%
  \let\@sf=\empty \ifhmode\edef\@sf{\spacefactor\the\spacefactor}\/\fi
  #1\insert\footins\bgroup
    \eightpoint\b@ls{10pt}\rm
    \interlinepenalty\interfootnotelinepenalty
%    \splittopskip\ht\strutbox % top baseline for broken footnotes
    \splitmaxdepth\dp\strutbox \floatingpenalty\@MM
    \leftskip\z@skip \rightskip\z@skip \spaceskip\z@skip \xspaceskip\z@skip
    \noindent\footstrut #1$\,$#2\strut\par
  \egroup
  \@sf\relax}

% Landscape

\def\landscape{%
  \global\TEMPDIMEN=\PageWidth
  \global\PageWidth=\PageHeight
  \global\PageHeight=\TEMPDIMEN
  \global\let\landscape=\relax%         stop them using \landscape again.
  \onecolumn
  \message{(landscape)}%
  \raggedbottom
}

%-------------------------------------------------------%
%               TWO COLUMN OUTPUT ROUTINE               %
%-------------------------------------------------------%

% Very slight redefinition of the \output routine of mn.tex, to allow footnotes.
\output{%
  \ifLeftCOL
    \global\setbox\LeftBOX=\vbox to \ZoneBSize{\box255\unvbox\ZoneBBOX
      \ifvoid\footins\else
        \vskip\skip\footins\unvbox\footins\fi
    }%
    \global\LeftCOLfalse
    \MakeRightCol
  \else
    \setbox\RightBOX=\vbox to \ZoneBSize{\box255\unvbox\ZoneBBOX
      \ifvoid\footins\else
        \vskip\skip\footins\unvbox\footins\fi
    }%
    \setbox\MidBOX=\hbox{\box\LeftBOX\hskip\ColumnGap\box\RightBOX}%
    \setbox\PageBOX=\vbox to \PageHeight{%
      \UnloadZoneA\box\MidBOX\UnloadZoneC}%
    \shipout\vbox{\PageHead\vbox to \PageHeight{\box\PageBOX\vss}\PageFoot}%
    \advancepageno
    \ifplate@page
      \shipout\vbox{%
        \sp@pagetrue
        \def\sp@type{plate}%
        \global\plate@pagefalse
        \PageHead\vbox to \PageHeight{\unvbox\plt@box\vfil}\PageFoot%
      }%
      \message{[plate]}%
      \advancepageno
    \fi
    \global\LeftCOLtrue
    \CleanStack
    \MakePage
  \fi
}

% Startup message

\Warn{\start@mess}

\newif\ifCUPmtplainloaded % for use in documents
\ifprod@font
  \global\CUPmtplainloadedtrue
\fi

\def\mnmacrosloaded{} % so articles can see if a format file has been used.

\catcode `\@=12 % @ signs are non-letters

% \dump

% end of mn.tex

%% file: shadowing.bbl
\section*{References}
\beginrefs
\bibitem Bennett J. C., Anderson A. P., McInnes P. A., Whitaker A. J. T.,
	1976, IEEE Trans. Antennas. Propagat., 24, 295
\bibitem Birkinshaw M., 1990, in Mandolesi N., Vittorio N., eds,
	The Cosmic Microwave Background: 25 Years Later, 
	Kluwer, Dordrecht, p. 77
\bibitem Carlstrom J. E. et al., 2000, Phys. Scripta, T85, 148
\bibitem Church S. E., 1995, MNRAS, 272, 551
\bibitem Hansen R. C., 1989, Proc. IEEE, 77, 659
\bibitem Jones M. et al., 1993, Nat, 365, 320
\bibitem Lay O. P., Halverson N. W., 2000, ApJ, 543, 787
\bibitem Padin S., Cartwright J. K., Joy M., Meitzler J. C., 2000, 
	IEEE Trans. Antennas Propagat., 48, 836
\bibitem Ryle M., 1962, Nat, 194, 517
\bibitem Ryle M., Hewish A., Shakeshaft J. R., 1959, IRE Trans. Antennas
	Propagat., 7, 120
\bibitem Scott P. F., Ryle M., 1977, MNRAS, 178, 539
\bibitem Subrahmanyan R., 2002, in 
	Chen L. -W., Ma C. -P., Ng K. -W., Pen U. -L., eds,
	ASP Conf. Ser. Vol. 257,
	AMiBA 2001: High-z clusters, missing baryons, and CMB polarization.
	Astron. Soc. Pac., San Fransisco, p. 309 
\bibitem Sunyaev R. A., Zeldovich Ya. B., 1972, Comm. Astrophys. Sp. Phys.,
	4, 173
\endrefs
